\documentclass[prd,aps,preprint,nofootinbib,tightenlines,showpacs]{revtex4-1}
\usepackage{epsfig}
\usepackage{amsmath}
\usepackage{amsthm}
\usepackage{amssymb}
\vspace{1in}

\usepackage{color}

\newcommand{\la}{\langle}
\newcommand{\ra}{\rangle}

\begin{document}

\title{Sivers and Boer-Mulders functions in Light-Cone Quark Models}
\author{Barbara Pasquini}
\email{pasquini@pv.infn.it}
\affiliation{Dipartimento di Fisica Nucleare e Teorica, Universit\`{a} degli Studi di Pavia,}
\affiliation{
Istituto Nazionale di Fisica Nucleare, Sezione di Pavia, I-27100 Pavia, Italy}
\author{Feng Yuan}
\email{fyuan@lbl.gov}
\affiliation{Nuclear Science Division, Lawrence Berkeley National Laboratory,Berkeley, CA 94720}
\affiliation{RIKEN BNL Research Center, Building 510A, Brookhaven National Laboratory, Upton, NY 11973}

\pacs{12.39.ki,13.85.Qk,13.88.+e}
\vspace{0.5in}
\begin{abstract}
Results for the naive-time-reversal-odd quark distributions
in a light-cone quark model are presented.
The final-state interaction
effects are generated via single-gluon exchange mechanism.
The formalism of light-cone wave functions is used to derive general
expressions in terms of overlap of wave-function amplitudes
describing the different orbital angular momentum components of the nucleon.
In particular, the model predictions show a dominant contribution from
$S$- and $P$-wave interference in the Sivers function and a significant
contribution also from the interference of $P$ and $D$ waves in the
Boer-Mulders function. The favourable comparison with existing
phenomenological parametrizations motivates further  applications
to describe azimuthal asymmetries in hadronic reactions.
\end{abstract}

\maketitle
\def \p {\partial}
\def \dd {\phi_{u\bar dg}}
\def \ddp {\phi_{u\bar dgg}}
\def \pq {\phi_{u\bar d\bar uu}}
\def \jpsi {J/\phi}
\def \phip {\phi^\prime}
\def \to {\rightarrow}
\def \abst {\vert t \vert}
\def\bfsig{\mbox{\boldmath$\sigma$}}
\def\DT{\mbox{\boldmath$\Delta_T$}}
\def \jpsi {J/\phi}
\def\bfej{\mbox{\boldmath$\varepsilon$}}
\def\FT{T_R \vert_{t\to 0}}
\def\FV{T_V \vert_{t\to 0}}
\newcommand{\bea}{\begin{eqnarray}}
\newcommand{\eea}{\end{eqnarray}}
\newcommand{\be}{\begin{equation}}
\newcommand{\ee}{\end{equation}}
\newcommand{\nn}{\nonumber}
\def\ket#1{\hbox{$\vert #1\rangle$}}   
\def\bra#1{\hbox{$\langle #1\vert$}}   

\section{Introduction}

Transverse momentum dependent parton distributions (TMDs), as an
important extension to the usual Feynman parton distributions,
have attracted much attention in hadronic physics from both
experiment and theory sides. Various hadronic processes
have been used and proposed to study
these distributions~\cite{
Cahn:1978se,Collins:1984kg,Sivers:1989cc,Efremov:1992pe,Collins:1992kk,Collins:1993kq,Kotzinian:1994dv,Mulders:1995dh,Boer:1997nt,Boer:1997mf,Boer:1999mm,Brodsky:2002cx,Collins:2002kn,Belitsky:2002sm,Bacchetta:2004zf,Cherednikov:2007tw,D'Alesio:2007jt,Barone:2001sp,Goeke:2005hb,Bacchetta:2006tn,Vogelsang:2005cs}.
Together with the generalized parton distributions (GPDs)
(for reviews, see~\cite{Goeke:2001tz,Diehl:2003ny,Ji:2004gf,Belitsky:2005qn,Boffi:2007yc}),
TMDs  shall lead us
to a comprehensive picture of parton distributions
inside the nucleon, in particular, in a three-dimension fashion.

Phenomenologically, in order to extract these distribution functions
from experiments, we have to ensure that the QCD factorization
applies in the associated processes. These issues have been
extensively discussed in the last few years, and the relevant factorization
theorem has been built up for a number of semi-inclusive processes,
such as semi-inclusive hadron production in deep inelastic
scattering and low transverse momentum Drell-Yan lepton pair
production in hadronic collisions~\cite{Collins:1981uk,Ji:2004wu,Collins:2004nx}. In the last few years, there has also been a remarkable experimental progress on
experimental measurements (see Ref.~\cite{D'Alesio:2007jt} and references therein).
More importantly, the proposed future experiments shall provide
more constraints on these distribution functions.

Meanwhile, reasonable model calculations of these transverse
momentum dependent parton distributions have been proposed~\cite{Jakob:1997wg,Efremov:2002qh,Efremov:2003eq,Yuan:2003wk,Pobylitsa:2003ty,Efremov:2004qs,Efremov:2004tp,Burkardt:2003uw,Collins:2005ie,Collins:2005rq,Efremov:2006qm,Kotzinian:2006dw,Brodsky:2006hj,Pasquini:2006iv,Meissner:2007rx,Anselmino:2007fs,Anselmino:2008jk,Gamberg:2007wm,Gamberg:2003ey,Bacchetta:2007wc,Avakian:2007mv,Avakian:2007xa,Pasquini:2008ax,Bacchetta:2008af,Courtoy:2008vi,Courtoy:2008dn,Courtoy:2009pc,Avakian:2008dz,Anselmino:2008sga,Arnold:2008ap,Efremov:2009ze,She:2009jq,Meissner:2009ww,Gamberg:2009uk,Bianconi:2006yq}.
These calculations promoted our understanding
of the nucleon structure, and have been playing very important
role as a first step to describe the experimental observations
of the associated phenomena. In particular, these models
provide us an intuitive way to connect the physical observables
and the key input for the nucleon structure model, such
as the quark spin and orbital angular momentum contributions
to the proton spin.

Transverse momentum dependent quark distributions are defined
through the following quark-density matrix
\begin{equation}
{\cal M}(x,\boldsymbol k_\perp)=\int\frac{d\xi^-d^2\boldsymbol \xi_\perp}{(2\pi)^3}e^{-ik\cdot \xi}
\langle PS|\bar\psi(\xi){\cal L}^\dagger_\xi{\cal L}_0\psi(0)|PS\rangle \ ,
\end{equation}
where $x$ and $\boldsymbol k_\perp$ are the longitudinal momentum fraction
and transverse momentum carried by the quark, respectively.
Nucleon's momentum $P$ is dominated by the plus component
$P^+=(P^0+P^z)/\sqrt{2}$, and $S$ represents the polarization
vector.
 In the above equation, the gauge link ${\cal L}$ is
very important to retain the gauge invariance and leading to
nonzero naive-time-reversal-odd (T-odd) quark distributions. Among the leading order eight quark
TMDs, six of them are called the naive-time-reversal
even (T-even), whereas the rest two belong to the T-odd
distributions. One is the so-called quark Sivers function, which
describes the quark transverse momentum distribution correlated to
the transverse polarization vector of the nucleon. The other is
the so-called Boer-Mulders function, and usually interpreted
as the transverse momentum correlated with the quark transverse
polarization. Both quark distributions contribute to the
azimuthal asymmetries in hadronic reaction processes.

In Ref.~\cite{Pasquini:2008ax}, we have calculated the
T-even quark distributions in a light-cone quark model, extending
previous works on the parton distribution functions
(PDFs)~\cite{Pasquini:2006iv}, the GPDs~\cite{Boffi:2002yy,Boffi:2003yj,Pasquini:2005dk,Pasquini:2007xz},
 nucleon form factors~\cite{Pasquini:2007iz} and distribution amplitudes~\cite{Pasquini:2009ki}.
Such a model, based on the light-cone wave-function (LCWF)
overlap representation, is able to capture
the relevant information on the three-quark contribution to different observables.
These calculations are well suited to illustrate the relevance
of the different orbital angular
momentum components of the nucleon wave function, and
provide an intuitive picture for the physical
meaning of the quark TMDs. Moreover, they can
be regarded as initial input for  phenomenological studies
for the semi-inclusive processes where quark TMDs
play a very important role~\cite{Boffi:2009sh}.

In this paper, we extend these works  to the T-odd
quark distributions.
The unique feature for the latter distributions is the final/initial
state interaction effects. Without these effects, the T-odd parton
distributions would vanish. In the model calculation, these
interactions are calculated by taking into account the one-gluon
exchange mechanism between the struck quark and the nucleon spectators
described by (real) LCWFs.
This approach is complementary to a recent work~\cite{Brodsky:2010vs} where
the rescattering effects are incorporated in
augmented LCWFs containing an imaginary phase
which depends on the choice of advanced or retarded boundary condition for the gauge potential in the light-cone gauge.
Recently, there has also been interesting
study to go beyond the one-gluon exchange approximation, by resumming
 all order contributions~\cite{Gamberg:2009uk,Burkardt:2003uw}.

The rest of the paper is organized as follows. In Sec.~II, we briefly
introduce the light-cone quark model, explaining its physical
content and giving results for the light-cone wave-function amplitudes
describing the different orbital angular momentum components of the
nucleon state.
In Sec.~III, we derive the quark
Sivers function. We present a general formalism in terms of overlap of
light-cone wave-function amplitudes, and then apply it to a specific
light-cone quark model wave function. The corresponding formalism
for the Boer-Mulders function is described  in Sec.~IV.
 The model results  for the T-odd distributions are
presented in Sec.~V
and compared to different phenomenological parametrizations.
Finally, we conclude with a section summarizing our findings.

\section{Light-Cone amplitudes in a constituent quark model}

\label{sect:lcwf}

The wave-function amplitudes in light-cone quantization for the three-quark
Fock state of the nucleon has been studied extensively in the literature~\cite{Brodsky:1997de}.
According to the total quark orbital angular momentum projection,
these wave-function amplitudes are classified into $l_z=0$, $l_z=1$, $l_z=2$,
$l_z=-1$ components for total spin $+1/2$ of nucleon, i.e.,
\begin{eqnarray}
|P\uparrow\rangle_{uud}=|P\uparrow\rangle^{l_z=0}_{uud}
+|P\uparrow\rangle^{l_z=1}_{uud}
+|P\uparrow\rangle^{l_z=-1}_{uud}
+|P\uparrow\rangle^{l_z=2}_{uud}.\label{eq:1}
\end{eqnarray}
For completeness, we list the parametrization for these
wave-function amplitudes following Refs.~\cite{Ji:2002xn,Burkardt:2002uc,Ji:2003yj}:
\begin{eqnarray}
  |P\uparrow\rangle_{uud}^{l_z=0} &=&
\int d[1]d[2]d[3]\left( \psi_{uud}^{(1)}(1,2,3)
         + i\epsilon^{\alpha\beta}k_{1\alpha}k_{2\beta}  \psi_{uud}^{(2)}(1,2,3)\right) \nonumber \\
         &&  \times  \frac{\epsilon^{ijk}}{\sqrt{6}} b^{\dagger\, u}_{i\uparrow}(1)
             \left(b^{\dagger\, u}_{j\downarrow}(2)b^{\dagger\, d}_{k\uparrow}(3)
            -b^{\dagger\, d}_{j\downarrow}(2)b^{\dagger\, u}_{k\uparrow}(3)\right)
         |0\rangle \ , \label{lca1}\\
  |P\uparrow\rangle_{uud}^{l_z=1} &=& \int d[1]d[2]d[3]\left(k_{1\perp}^+
           \psi_{uud}^{(3)}(1,2,3)
         + k_{2\perp}^+ \psi_{uud}^{(4)}(1,2,3)\right) \nonumber \\
         &&  \times  \frac{\epsilon^{ijk}}{\sqrt{6}} \left( b^{\dagger\, u}_{i\uparrow}(1)
            b^{\dagger\, u}_{j\downarrow}(2)b^{\dagger\, d}_{k\downarrow}(3)
            -b^{\dagger\, d}_{i\uparrow}(1)b^{\dagger\, u}_{j\downarrow}(2)
             b^{\dagger\, u}_{k\downarrow}(3)\right)
         |0\rangle \ ,\label{lca2}
\end{eqnarray}
\begin{eqnarray}
  |P\uparrow\rangle_{uud}^{l_z=-1} &=& \int d[1]d[2]d[3]~k_{2\perp}^-
          \psi_{uud}^{(5)}(1,2,3) \nonumber \\
         &&  \times  \frac{\epsilon^{ijk}}{\sqrt{6}} b^{\dagger\, u}_{i\uparrow}(1)
             \left(
     b^{\dagger\, u}_{j\uparrow}(2)b^{\dagger\,d}_{k\uparrow}(3)
    -b^{\dagger\, d}_{j\uparrow}(2)b^{\dagger\, u}_{k\uparrow}(3)
            \right)
         |0\rangle \ ,\label{lca3}\\
  |P\uparrow\rangle_{uud}^{l_z=2} &=& \int d[1]d[2]d[3]~k_{1\perp}^+k_{3\perp}^+
         \psi_{uud}^{(6)}(1,2,3)  \nonumber \\
         &&  \times  \frac{\epsilon^{ijk}}{\sqrt{6}}
         b^{\dagger\, u}_{i\downarrow}(1)
         \left(b^{\dagger\, d}_{j\downarrow}(2)b^{\dagger\, u}_{k\downarrow}(3)
         -b^{\dagger\, u}_{j\downarrow}(2)d^
{\dagger}_{k\downarrow}(3)
            \right)
         |0\rangle \ ,\label{lca4}
\end{eqnarray}
 where $\alpha,\beta=1,2$ are transverse indexes and $k^\pm_{i\perp}=k^x_i\pm
k^y_i$.
In Eqs.~(\ref{lca1})-(\ref{lca4})
the integration measures are defined as
\begin{equation}
\label{eq:7}
d[1]d[2]d[3]=
\frac{dx_1dx_2dx_3}{\sqrt{x_1x_2x_3}}\delta\left(1-\sum_{i=1}^3 x_i\right)
\frac{d^2 \boldsymbol{k}_{1\perp}d^2\boldsymbol{k}_{2\perp}
d^2\boldsymbol{k}_{3\perp}}{[2(2\pi^3)]^2}
\delta\left(\sum_{i=1}^3 \boldsymbol{k}_{i\perp}\right),
\end{equation}
with $x_i$  the fraction of the longitudinal nucleon momentum carried by the quarks, and
$\boldsymbol{k}_{i\perp}$  their transverse momenta.
Furthermore, $b^{\dagger\, q}_{i,\,\lambda}$ and $b^q_{i,\,\lambda}$
are creation and annihilation operators of a quark with flavour $q$, helicity $\lambda$ and color $i$, respectively.
In the following, we will describe the above light-cone wave-function
amplitudes in a light-cone constituent quark model
(CQM) following Ref.~\cite{Pasquini:2008ax}.
Working in the so-called ``uds'' basis~\cite{Franklin:68,Capstick:1986bm}
the proton state is given in terms of a completely symmetrized wave function of the form
\begin{equation}
|P\uparrow\rangle=|P\uparrow\rangle_{uud}+|P\uparrow\rangle_{udu}+
|P\uparrow\rangle_{duu} \,.
\label{eq:2}
\end{equation}
In this symmetrization, the state $|P\uparrow\rangle_{udu}$ is obtained from
$|P\uparrow\rangle_{uud}$ by interchanging the second and third spin and space coordinates as well as the indicated quark type, with a similar interchange of the first and third coordinates for
$|P\uparrow\rangle_{duu}$.

Following the derivation outlined in Ref.~\cite{Boffi:2002yy}, we find that the $uud$ component of the light-cone state of the proton can be written as
\be\label{eq:12}
\ket{P,\Lambda}_{uud} = \sum_{\lambda_i,c_i}
\int d[1]d[2]d[3]
\Psi^{\Lambda,[f]}_{uud}(\{x_i,\boldsymbol{ k}_{i\perp };\lambda_i\})
\frac{\epsilon^{ijk}}{\sqrt{6}}
b^{\dagger\,u}_{i,\,\lambda_1}(1)
b^{\dagger\,u}_{j,\,\lambda_2}(2)
b^{\dagger\,d}_{k,\,\lambda_3}(3)
|0\rangle\, .
\ee
In Eq.~(\ref{eq:12}), assuming SU(6) spin-flavor symmetry, we can factorize
the LCWF $\Psi^{\Lambda,[f]}_{uud}(\{x_i,\boldsymbol{ k}_{i\perp };\lambda_i\})$
in a momentum-dependent  wave function and a spin-dependent part, i.e.,
\begin{eqnarray}
\label{eq:13}
   \Psi^{\Lambda,[f]}_{uud}(\{x_i,\boldsymbol{ k}_{i\perp }; \lambda_i\})
&=&
\tilde \psi(\{x_i,\boldsymbol{ k}_{i\perp }\})
\frac{1}{\sqrt{3}}\tilde\Phi_{\Lambda}(\lambda_1,\lambda_2,\lambda_3).
\end{eqnarray}
In the above equation the momentum-dependent function is given by
\begin{eqnarray}\label{eq:14}
\tilde \psi(\{x_i,\boldsymbol{ k}_{i\perp }\})=
2(2\pi)^3\bigg[\frac{1}{M_0}\frac{\omega_1\omega_2\omega_3}{x_1x_2x_3}\bigg]^{1/2}\psi(\{x_i,\boldsymbol{ k}_{i\perp }\}),
\end{eqnarray}
where $\psi(\{x_i,\boldsymbol{ k}_{i\perp }\})$ is symmetric under exchange of the momenta of
any quark pairs and is spherically symmetric, $\omega_i$ is the free-quark energy, and $M_0=\sum_i\omega_i$ is the mass of the non-interacting three-quark system.
The spin-dependent part  in Eq.~(\ref{eq:13}) is given by
\begin{eqnarray}
\tilde\Phi_{\Lambda}(\lambda_1,\lambda_2,\lambda_3)
&=&\sum_{\mu_1\mu_2\mu_3}
\langle 1/2,\mu_1; 1/2, \mu_2|1, \mu_1+\mu_2 \rangle
\langle 1, \mu_1+\mu_2;1/2, \mu_3| 1/2, \Lambda\rangle\nn\\
&&\times
D_{\mu_1\lambda_1}^{1/2*}(R_{cf}(x_1,\boldsymbol{ k}_{1\perp }))
D_{\mu_2\lambda_2}^{1/2*}(R_{cf}(x_2,\boldsymbol{ k}_{2\perp }))
D_{\mu_3\lambda_3}^{1/2*}(R_{cf}(x_3,\boldsymbol{ k}_{3\perp })).
\label{eq:15}
\end{eqnarray}
In Eq.~(\ref{eq:15}), $D_{\lambda\mu}^{1/2}(R_{cf}(x,\boldsymbol{ k}_\perp))$ is
the matrix element of the Melosh rotation $R_{cf}$~\cite{Melosh:74}
\begin{eqnarray}
D_{\lambda\mu}^{1/2}(R_{cf}(x,\boldsymbol{ k}_\perp)) &=&
\langle\lambda|R_{cf}(x,\boldsymbol{k}_\perp)|\mu\rangle\nonumber\\
&=&
\langle\lambda|\frac{m + xM_0 -
i\boldsymbol{\sigma}\cdot(\hat{\boldsymbol{z}}\times\boldsymbol{k}_\perp)}{\sqrt{(m
+ xM_0)^2 + \boldsymbol{k}^{\, 2}_\perp}}|\mu\rangle.
\label{eq:16}
\end{eqnarray}
The Melosh rotation corresponds to the unitary transformation which converts
the instant-form spin eigenstates (the Pauli spinors)
 to  light-front helicity eigenstates.
In particular, the light-cone spin wave function of Eq.~(\ref{eq:15})
is obtained from the transformation of the canonical-spin wave function with zero orbital angular momentum component.
The effects of the Melosh transformation
are immediately evident in the presence of the spin-flip term
$i\boldsymbol{\sigma}\cdot(\hat{\boldsymbol{z}}\times\boldsymbol{k}_\perp)$ in Eq.~(\ref{eq:16}).
Such a term generates non-zero orbital angular momentum, even if the original (instant-form) wave function only contained S-wave components.
Therefore, as a consequence of total angular momentum conservation, the LCWF
has components with
total quark helicity different from the nucleon helicity.
Making explicit the dependence on the quark helicities, the light-cone spin wave function of Eq.~(\ref{eq:15}) takes the following values:
\begin{eqnarray}
\label{eq:17}
\tilde \Phi_\uparrow\left(\uparrow,\uparrow,\downarrow\right)
&=&\prod_i\frac{1}{\sqrt{N(x_i,\boldsymbol{k}_{i\perp })}}
\frac{1}{\sqrt{6}}(2a_1a_2a_3+a_1 k_2^-k_3^+ +a_2k_1^-k_3^+),
\\\label{eq:18}
\tilde \Phi_\uparrow\left(\uparrow,\downarrow,\uparrow\right)
&=&\prod_i\frac{1}{\sqrt{N(x_i,\boldsymbol{ k}_{i\perp })}}
\frac{1}{\sqrt{6}}(-a_1a_2a_3+a_3 k_1^- k_2^+-2a_1k_2^+k_3^-),
\\\label{eq:19}
\tilde \Phi_\uparrow\left(\downarrow,\uparrow,\uparrow\right)
&=&\prod_i\frac{1}{\sqrt{N(x_i,\boldsymbol{ k}_{i\perp })}}
\frac{1}{\sqrt{6}}(-a_1a_2a_3+a_3 k_1^+k_2^--2a_2k_1^+k_3^-),
\\\label{eq:20}
\tilde \Phi_\uparrow\left(\uparrow,\downarrow,\downarrow\right)
&=&\prod_i\frac{1}{\sqrt{N(x_i,\boldsymbol{ k}_{i\perp })}}
\frac{1}{\sqrt{6}}(a_1a_2k_3^+-  k_1^- k_2^+k_3^+-2a_1 a_3k_2^+),
\\\label{eq:21}
\tilde \Phi_\uparrow\left(\downarrow,\uparrow,\downarrow\right)
&=&\prod_i\frac{1}{\sqrt{N(x_i,\boldsymbol{ k}_{i\perp })}}
\frac{1}{\sqrt{6}}(-k_1^+k_2^-k_3^++a_1a_2  k_3^+ -2a_2 a_3k_1^+),
\\\label{eq:21a}
\tilde \Phi_\uparrow\left(\downarrow, \downarrow, \uparrow\right)
&=&\prod_i\frac{1}{\sqrt{N(x_i,\boldsymbol{ k}_{i\perp })}}
\frac{1}{\sqrt{6}}(a_2 a_3 k_1^+ +a_1 a_3k_2^+ +2k_1^+k_2^+k_3^-),
\\\label{eq:22}
\tilde \Phi_\uparrow\left(\uparrow,\uparrow,\uparrow\right)
&=&\prod_i\frac{1}{\sqrt{N(x_i,\boldsymbol{ k}_{i\perp })}}
\frac{1}{\sqrt{6}}(-a_1a_3k_2^- - a_2a_3 k_1^-+2a_1a_2k_3^-),
\\\label{eq:23}
\tilde \Phi_\uparrow\left(\downarrow,\downarrow,\downarrow\right)
&=&\prod_i\frac{1}{\sqrt{N(x_i,\boldsymbol{ k}_{i\perp })}}
\frac{1}{\sqrt{6}}(-a_2k_1^+k_3^+ - a_1 k_2^+k_3^++2a_3k_1^+k_2^+),
\end{eqnarray}
where $a_i=(m+x_i M_0)$, and $N(x_i,\boldsymbol{ k}_{i\perp })=
[(m+x_i M_0)^2+ \boldsymbol{ k}^2_{i\perp}]$.

Taking into account the quark-helicity dependence in Eqs.~(\ref{eq:17})-(\ref{eq:23}), the nucleon state can be mapped out into the different angular momentum components. After straightforward algebra, one finds the following representation for the nucleon wave-function amplitudes in the light-cone CQM
\begin{eqnarray}
\psi^{(1)}(1,2,3)&=&\tilde \psi(\{x_i,\boldsymbol{ k}_{i\perp }\})\nn\\
&&\times
\prod_i\frac{1}{\sqrt{N(x_i,\boldsymbol{ k}_{i\perp })}}
\frac{1}{\sqrt{3}}(
-a_1 a_2 a_3
+a_3 \boldsymbol{ k}_{1\perp}\cdot \boldsymbol{ k}_{2\perp}
+2a_1 \boldsymbol{ k}_{1\perp}\cdot \boldsymbol{ k}_{2\perp}
+2a_1 \boldsymbol{ k}_{2\perp}^2),\nn\\
&&\label{eq:24}\\
\psi^{(2)}(1,2,3)&=&\tilde \psi(\{x_i,\boldsymbol{ k}_{i\perp }\})
\prod_i\frac{1}{\sqrt{N(x_i,\boldsymbol{ k}_{i\perp })}}\frac{1}{\sqrt{3}}
(a_3 + 2 a_1),
\label{eq:25}
\end{eqnarray}
\begin{eqnarray}
\psi^{(3)}(1,2,3)&=&-\tilde \psi(\{x_i,\boldsymbol{ k}_{i\perp }\})
\prod_i\frac{1}{\sqrt{N(x_i,\boldsymbol{ k}_{i\perp })}}
\frac{1}{\sqrt{3}}(a_1 a_2 + \boldsymbol{ k}_{2\perp}^2),
\label{eq:26}\\
\psi^{(4)}(1,2,3)&=&-\tilde \psi(\{x_i,\boldsymbol{ k}_{i\perp }\})
\prod_i\frac{1}{\sqrt{N(x_i,\boldsymbol{ k}_{i\perp })}}
\frac{1}{\sqrt{3}}(a_1 a_2 +  2a_3 a_1-\boldsymbol{ k}_{1\perp}^2 -2
\boldsymbol{ k}_{1\perp}\cdot \boldsymbol{ k}_{2\perp}),
\label{eq:27}\\
\psi^{(5)}(1,2,3)&=&\tilde \psi(\{x_i,\boldsymbol{ k}_{i\perp }\})
\prod_i\frac{1}{\sqrt{N(x_i,\boldsymbol{ k}_{i\perp })}}
\frac{1}{\sqrt{3}}(a_1 a_3),
\label{eq:28}\\
\psi^{(6)}(1,2,3)&=&\tilde \psi(\{x_i,\boldsymbol{ k}_{i\perp }\})
\prod_i\frac{1}{\sqrt{N(x_i,\boldsymbol{ k}_{i\perp })}}
\frac{1}{\sqrt{3}} a_2.
\label{eq:29}
\end{eqnarray}
Notice that the results in Eqs.~(\ref{eq:24})-(\ref{eq:29}) follow from the spin and orbital angular momentum structure generated from the Melosh rotations, and are independent on the functional form of the momentum-dependent wave function.

\section{Sivers function}
\label{sec:sivers}
The quark Sivers function can be calculated from the following
definition
\begin{equation}
\label{sivers}
f_{1T}^\perp(x,\boldsymbol k^{\, 2}_\perp)=-i(k^x+ik^y)\frac{M}{2\boldsymbol k^{\, 2}_\perp}
\int\frac{d\xi^-d^2\boldsymbol \xi_\perp}{(2\pi)^3}
e^{-i(\xi^- k^+-\boldsymbol\xi_\perp\cdot\boldsymbol k_\perp)}
\langle P\uparrow|\bar\psi(\xi^-,\boldsymbol\xi_\perp){\cal L}^\dagger_\xi\gamma^+{\cal L}_0\psi(0)|P
\downarrow\rangle \ .
\end{equation}
As we discussed in the Introduction, the gauge link is crucial
to obtain a non-zero Sivers function. In the covariant gauge, the gauge
link can be reduced to the light-cone gauge link\footnote{An off-light-cone
gauge link has to be used to regulate the light-cone singularities for
higher-order calculations. In this paper, we will not encounter this
singularity. Therefore, we will simply adopt the gauge link
along the light-cone direction in covariant gauge and the transverse
gauge link at spatial infinity in light-cone gauge.}. According to
the light-cone wave function model, in the following calculations we
choose the light-cone gauge $A^+=0$, where the gauge link reduces to
a transverse gauge link at $\xi^-=\infty$, i.e.,
\begin{equation}
{\cal L}_\xi|_{A^+=0}={\cal P}\exp\left(-ig\int_{\boldsymbol \xi_\perp}^\infty
{\rm d}^2\boldsymbol \zeta_\perp\cdot \boldsymbol A_\perp(\xi^-=\infty,\boldsymbol\zeta_\perp)\right) \ .
\end{equation}
In the Sivers function of Eq.~(\ref{sivers}),
we will expand the above gauge link to take into account the contribution
from the one-gluon exchange diagram.
Furthermore, in the light-cone gauge the
gluon propagator takes the following form
\begin{equation}
d^{\mu\nu}(q)=-g^{\mu\nu}+\frac{n^\mu q^\nu+n^\nu q^\mu}{[n\cdot q]} \ ,
\end{equation}
where $n$ is the light-like vector $n^2=0$ and $n\cdot q=q^+$.
The gluon propagator has a light-cone singularity, as can be
seen from the above equation.
We will adopt the principal-value prescription
to regulate this singularity. We have also checked that the final results
do not
depend on the prescription.\footnote{For example,
if we choose the so-called advanced
boundary condition for the gauge potential, the transverse gauge link becomes
unit, whereas the above gluon propagator generates phases
which allow to recover the previous results with the principal-value prescription.}
Under this prescription, there is no phase contribution from
the above propagator. However, the transverse gauge link expansion,
when combining with the $n^-\boldsymbol q_\perp/n\cdot q$ factor of the above equation,
leads to the following expression
\begin{equation}
\frac{e^{iq^+\infty}}{q^+}=i\pi\delta(q^+)\ .
\end{equation}
This contribution  provides the phase needed to
generate a non-zero Sivers function.
The dominance in the gluon propagator of the
$n^-\boldsymbol q_\perp/n\cdot q,$ with the $\perp$
index coming from the contraction with the transverse gauge link,
also simplifies the interactions between
the quark fields, since the  quark scattering conserves the helicity.

\begin{figure}
\centerline{
\epsfig{file=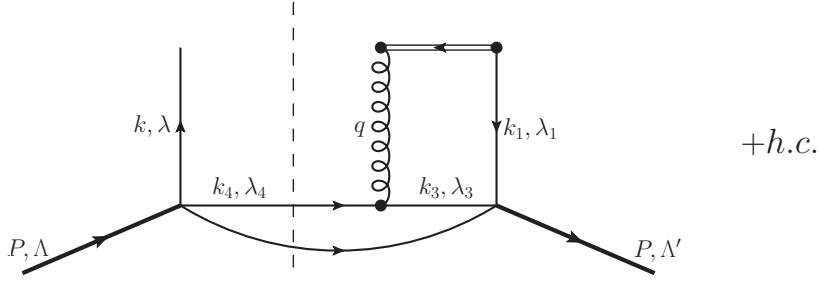,  width=0.6\columnwidth}}
\caption{\label{fig1}
The leading contribution from the one-gluon exchange mechanism to
the T-odd distribution functions.}
\end{figure}

Finally, we obtain the following expression for the
quark Sivers function
\begin{eqnarray}
&&f_{1T}^{\perp\,q}(x,\boldsymbol k^{\, 2}_\perp)
=-g^2\frac{k^x+ik^y}{\boldsymbol k^{\, 2}_\perp}\frac{M}{2}
\frac{1}{(2\pi)^{11}}\frac{1}{\sqrt{(2k^+)(2k^+_1)}}
\int\frac{{\rm d}k_3^+{\rm d}^2\,\boldsymbol k_{3\perp}}{\sqrt{(2k_3^+)(2k^+_4)}}
\int{\rm d}^2\, \boldsymbol q_\perp
 \nonumber\\
&&\times
\Big\{\frac{1}{\boldsymbol{q}^{\,2}_\perp}\sum_{\lambda_1,\lambda_3}
\sum_f\sum_{i.j}\sum_{k,l}
T^a_{ij}T^b_{kl}\delta_{ab}
\langle P\uparrow|b^{\dagger\, q}_{i\lambda_1}(k_1)b^q_{j\lambda_1}(k)
b^{\dagger\, f}_{k\lambda_3}(k_3)b^f_{l\lambda_3}(k_4)
|P\downarrow\rangle\Big\},
\label{eq:sivers1}
\end{eqnarray}
where  the quark momenta are  defined as $k_1=k-q$, $k_4=k_3-q$,
$T^a$ is the $SU_c(6)$ Gell-Mann matrix and $g$ is the gluon
coupling with the quark field. Equation~(\ref{eq:sivers1}) corresponds
to the diagrams in Fig.~\ref{fig1} with  $\lambda=\lambda_1$ and
$\lambda_4=\lambda_3$, for the helicity of the interacting and spectator quarks, respectively,  and $\Lambda=-\Lambda'$ for the helicity of the nucleon in the initial and final states.
A few comments are in order to explain the above derivations.
First, we have made an approximation for the interaction vertex
between the gauge field from the gauge link and the quark fields
in the proton wave function, by the covariant interaction form.
In principle, we shall use the light-cone time-order perturbation
theory to describe this interaction. However, we expect the modification
being beyond the approximation we  made in modelling the light-cone
wave function itself.
Nevertheless, it will be interesting to
check how large these effects would be.
Second, we used the perturbation theory to calculate the
final-state interaction effects. For numerical estimate,
we choose a reasonable value for the strong coupling constant (see Sec.~\ref{sect:results}).
Meanwhile, we notice it may be not appropriate to use
a perturbative coupling for this non-perturbative calculations.
We regarded this as an important theoretical uncertainty,
which exists in all model calculations of the Sivers function.

As we discussed, in Eq.~(\ref{eq:sivers1}) the quark helicity is
conserved. On the other side,
the hadron helicity flips from the initial to the final state.
As a consequence, non-zero results for the Sivers function
can be obtained only with a transfer of one unit
of orbital angular momentum between the initial and the final nucleon states.
\newline
\noindent
Inserting in Eq.~(\ref{eq:sivers1}) the
light-cone wave-function amplitude decomposition of the
nucleon state introduced  in Sec.~\ref{sect:lcwf}, one finds the following results in terms of the
amplitudes $\psi^{(i)}$
\begin{eqnarray}
f_{1T}^{\perp\,q}(x,\boldsymbol k_\perp^{\, 2})&=&-\frac{2}{3}g^2M\frac{k^x+ik^y}{\boldsymbol k^{\, 2}_\perp}
\int \frac{{\rm d^2}\, \boldsymbol q_\perp}{(2\pi)^2}\frac{1}{\boldsymbol{q}^{\,2}_\perp}
\int{\rm d}x'\int{\rm d}^2\boldsymbol t '_\perp
\int d[1]d[2]d[3] \sqrt{x_1 x_2 x_3}\,{\cal F}^{\perp\,q}.\nonumber\\
&&\label{eq:sivers-overlap}
\end{eqnarray}
The function ${\cal F}^{\perp\,q}$ for $u$ quark is given by
\begin{eqnarray}
{\cal F}^{\perp\, u}=
A^{(1,2)}\phi^{(3,4)}(1,2,3)-A^{(3,4)}\phi^{(1,2)}(1,2,3)-
	A^{(5)}\phi^{(6)}(1,2,3)+A^{(6)}\phi^{(5)}(1,2,3),
\label{eq:F-up}
\end{eqnarray}
where
\begin{eqnarray}
\phi^{(1,2)}(1,2,3)&=&\psi^{(1)}(1,2,3)-i(k_1^xk_2^y-k_1^yk_2^x)\psi^{(2)}(1,2,3),\nonumber\\
\phi^{(3,4)}(1,2,3)&=&k_1^-\psi^{(3)}(1,2,3)
	+k_2^- \psi^{(4)}(1,2,3) ,\nonumber\\
\phi^{(5)}(1,2,3)&=&k_2^+ \psi^{(5)}(1,2,3)
	-k_3^+ \psi^{(5)}(1,3,2) ,\nonumber\\
\phi^{(6)}(1,2,3)&=&k_1^-k_3^- \psi^{(6)}(1,2,3)
		-k_1^-k_2^- \psi^{(6)}(1,3,2) .
\end{eqnarray}
The functions $A$ in Eq.~(\ref{eq:F-up}) are defined through
\begin{eqnarray}
A^{(1,2)}&=&\delta^3(k-k_1)\Big[\delta^3(t'-k_2)\phi^{(1,2)*}(\hat 2,1'',3)
+\delta^3(t'-k_3)\phi^{(1,2)*}(2,1'',\hat 3)\Big]\nonumber\\
&&+
\delta^3(k-k_2)\Big[\delta^3(t'-k_1)\Big(2\phi^{(1,2)*}(2'',\hat 1,3)
+\phi^{(1,2)*}(3,\hat 1,2'')\Big)\nonumber\\
&&
+\delta^3(t'-k_3)\Big(2\phi^{(1,2)*}(2'',1,\hat 3)
+\phi^{(1,2)*}(\hat 3,1,2'')\Big)\Big]\nonumber\\
&&+\delta^3(k-k_3)\Big[\delta^3(t'-k_1)\Big(\phi^{(1,2)*}(2,\hat 1,3'')
                      +\phi^{(1,2)*}(3'',\hat 1,2)\Big)\nonumber\\
&&+\delta^3(t'-k_2)\Big(\phi^{(1,2)*}(\hat 2,1,3'')
+\phi^{(1,2)*}(3'',1,\hat 2)\Big)\Big] ,
\nonumber
\end{eqnarray}
\begin{eqnarray}
A^{(3,4)}&=&\delta^3(k-k_2)\Big[\delta^3(t'-k_1)\phi^{(3,4)*}(2'',\hat 1,3)+
\delta^3(t'-k_3)\phi^{(3,4)*}(2'', 1,\hat 3)\Big]\nonumber\\
&&+
\delta^3(k-k_1)\Big[
\delta^3(t'-k_2)\Big(2\phi^{(3,4)*}(\hat 2,1'',3)+\phi^{(3,4)*}(\hat 2,3,1'')\Big)\nonumber\\
&&+\delta^3(t'-k_3)\Big(2\phi^{(3,4)*}(2,1'',\hat 3)+\phi^{(3,4)*}(2,\hat 3,1'')\Big)]\Big]\nonumber\\
&&+\delta^3(k-k_3)\Big[\delta^3(t'-k_1)\Big(\phi^{(3,4)*}(2,\hat 1,3'')+
	 	\phi^{(3,4)*}(2,3'',\hat 1)\Big)\nonumber\\
&&+\delta^3(t'-k_2)\Big(\phi^{(3,4)*}(\hat 2,1,3'')+
	 	\phi^{(3,4)*}(\hat 2,3'',1)\Big)\Big],\nonumber
\end{eqnarray}
\begin{eqnarray}
A^{(5)}&=&\delta^3(k-k_1)\Big[\delta^3(t'-k_2)\Big(\phi^{(5)*}(1'',\hat 2,3)+
		\phi^{(5)*}(\hat 2,1'',3)\Big)\nonumber\\
&&+\delta^3(t'-k_3)\Big(\phi^{(5)*}(1'',2,\hat 3)+
		\phi^{(5)*}(2,1'',\hat 3)\Big)\Big]\nonumber\\
&&+\delta^3(k-k_2)\Big[\delta^3(t'-k_1)\Big(\phi^{(5)*}(\hat 1,2'',3)+
		\phi^{(5)*}(2'',\hat 1,3)\Big)\nonumber\\
&&+\delta^3(t'-k_3)\Big(\phi^{(5)*}(1,2'',\hat 3)+
		\phi^{(5)*}(2'',1,\hat 3)\Big)\Big],\nonumber
\end{eqnarray}
\begin{eqnarray}
A^{(6)}&=&\delta^3(k-k_1)\Big[\delta^3(t'-k_2)\Big(\phi^{(6)*}(1'',\hat 2,3)+
	\phi^{(6)*}(\hat 2,1'',3)\Big)\nonumber\\
&&+\delta^3(t'-k_3)\Big(\phi^{(6)*}(1'', 2,\hat 3)+
	\phi^{(6)*}( 2,1'',\hat 3)\Big)\Big]\nonumber\\
&&+\delta^3(k-k_2)\Big[\delta^3(t'-k_1)\Big(\phi^{(6)*}(\hat 1,2'',3)+
	\phi^{(6)*}(2'',\hat 1,3)\Big)\nonumber\\
&&+\delta^3(t'-k_3)\Big(\phi^{(6)*}(1,2'',\hat 3)+
	\phi^{(6)*}(2'',1,\hat 3)\Big)\Big],
\label{eq:A}
\end{eqnarray}
where the quark coordinates are
$\imath''=(x,\boldsymbol k_\perp -\boldsymbol q_\perp)$, and
$\hat{\imath}=(x',\boldsymbol{t}\,'_{\perp}+\boldsymbol{q}_\perp)$, $\delta^3(k-k_i)=\delta(x-x_i)\delta^2(\boldsymbol k_\perp-\boldsymbol k_{i\perp})$ and
we used the notation
$\delta^3(t'-k_i)=\delta(x'-x_i)\delta^2(\boldsymbol t\, '_\perp-\boldsymbol k_{i\perp})$.
In the above equations, the complex conjugate only acts on the wave function $ {\psi}^{(i)}$.

In Eq.~(\ref{eq:sivers-overlap}), the contributions from
the functions $A^{(1,2)}$ and $A^{(3,4)}$
describe the interference between $S$ and $P$ waves, while the terms with
 $A^{(5)}$ and $A^{(6)}$  correspond to the contribution from
$P-D$ wave interference.

Similarly for the d-quark, one has
\begin{eqnarray}
{\cal F}^{\perp\, d}=
B^{(1,2)}\phi^{(3,4)}(1,2,3)-B^{(3,4)}\phi^{(1,2)}(1,2,3)-
	B^{(5)}\phi^{(6)}(1,2,3)+B^{(6)}\phi^{(5)}(1,2,3) ,
\label{eq:F-down}
\end{eqnarray}
where the terms with $B^{(1,2)}$ and $B^{(3,4)}$
describe the interference between $S$ and $P$ waves, while the terms with
 $B^{(5)}$ and $B^{(6)}$  correspond to the contribution from
$P-D$ wave interference.
The explicit expression for these functions is
\begin{eqnarray}
B^{(1,2)}&=&\delta^3(k-k_3)\Big[\delta^3(t'-k_2)\phi^{(1,2)*}(\hat 2,1,3'')
+\delta^3(t'-k_1)\phi^{(1,2)*}(2,\hat 1,3'')\Big]\nonumber\\
&&+\delta^3(k-k_1)
\Big[\delta^3(t'-k_2)
\Big(\phi^{(1,2)*}(\hat 2,1'',3)+\phi^{(1,2)*}(3,1'',\hat 2)\Big)\nonumber\\
&&+\delta^3(t'-k_3)
\Big(\phi^{(1,2)*}(2,1'',\hat 3)+\phi^{(1,2)*}(\hat 3,1'',2)\Big)\Big],
\nonumber
\end{eqnarray}
\begin{eqnarray}
B^{(3,4)}&=&\delta^3(k-k_3)\Big[\delta^3(t'-k_1)
\phi^{(3,4)*}(2,\hat 1,3'')+\delta^3(t'-k_2)
\phi^{(3,4)*}(\hat 2,1,3'')\Big]\nonumber\\
&&
+\delta^3(k-k_2)\Big[\delta^3(t'-k_1)\Big(\phi^{(3,4)*}(2'',\hat 1,3)
+\phi^{(3,4)*}(2'',3,\hat 1)\Big)\nonumber\\
&&+\delta^3(t'-k_3)\Big(\phi^{(3,4)*}(2'',1,\hat 3)
+\phi^{(3,4)*}(2'',\hat 3,1)\Big)\Big],\nonumber\\
&&\nonumber\\
B^{(5)}&=&\delta^3(k-k_3)\Big[\delta(t'-k_2)\Big(\phi^{(5)*}(1,\hat 2,3'')+
		\phi^{(5)*}(\hat 2,1,3'')\Big)\nonumber\\
&&+\delta(t'-k_1)\Big(\phi^{(5)*}(\hat 1,2,3'')+
		\phi^{(5)*}(2,\hat 1,3'')\Big)\Big] ,\nonumber\\
&&\nonumber\\
B^{(6)}&=&\delta^3(k-k_3)\Big[\delta^3(t'-k_1)\Big(\phi^{(6)*}(\hat 1,2,3'')+
	\phi^{(6)*}(2,\hat 1,3'')\Big)\nonumber\\
&&+\delta^3(t'-k_2)\Big(\phi^{(6)*}(1,\hat 2,3'')+
	\phi^{(6)*}(\hat 2,1,3'')\Big)\Big].
\label{eq:B}
\end{eqnarray}
In the above equations, the complex conjugate only acts on the wave function $ {\psi}^{(i)}$.

Using the CQM expressions for the three-quark light cone amplitudes given in Sec.~\ref{sect:lcwf}, we obtain the following results
for the Sivers function
\begin{eqnarray}
&&f_{1T}^{\perp\,q}(x,\boldsymbol k^{\, 2}_\perp)=-\frac{2}{3}g^2M
\frac{k^x+ik^y}{\boldsymbol k^{\,2}_\perp}
\int \frac{{\rm d}^2\boldsymbol q_\perp}{(2\pi)^2}\frac{1}{\boldsymbol q^{\; 2}_\perp}
\int{\rm d}x'\int{\rm d}^2\boldsymbol t'_\perp
\int
d[1]d[2]d[3] \sqrt{x_1 x_2 x_3}\nonumber\\
&&\times \delta(x-x_3)
\delta^2(\boldsymbol k_\perp-\boldsymbol k_{3\perp})\delta(x'-x_1)
\delta^2(\boldsymbol t\, '_\perp-\boldsymbol k_{1\perp})
\,\psi^*(\{x'_i\},\{\boldsymbol k\,'_{i\perp}\})
\,\psi(\{x_i\},\{\boldsymbol k_{i\perp}\})\nonumber\\
&&\times3\delta_{\tau_3\tau_q}\
\left\{
\delta_{\tau_q 1/2}X^{00}(\{\boldsymbol k\,'_i\},\{\boldsymbol k_i\})
+\frac{1}{3}[\delta_{\tau_q 1/2}+2\delta_{\tau_q -1/2}]
X^{11}(\{\boldsymbol k\,'_i\},\{\boldsymbol k_i\})\right\},\label{eq:sivers-lcwf}
\end{eqnarray}
where the quark momenta in the final state are $(x'_3=x,\boldsymbol k\, '_{3\perp}=\boldsymbol k_{3\perp}-\boldsymbol q_\perp)$,
$(x'_1=x',\boldsymbol k\,'_1=\boldsymbol t\, '_\perp+\boldsymbol q_\perp)$, $(x'_2=x_2,\boldsymbol k\, '_{2\perp}=\boldsymbol k_{2\perp})$.
In Eq.~(\ref{eq:sivers-lcwf}), the functions $X^{00}$ and $X^{11}$ are given by
\begin{eqnarray}
X^{00}(\{\boldsymbol k\,'_i\},\{\boldsymbol k_i\})
&=& \prod_{i=1}^3 N^{-1}(\boldsymbol{k}\,'_i) N^{-1}(\boldsymbol{k}_i)
(i\,B_{3x}+B_{3y}) (A_1A_2 +
\boldsymbol{B}_1\cdot\boldsymbol{B}_2),\label{eq:x00_flip}\\
X^{11}(\{\boldsymbol k\,'_i\},\{\boldsymbol k_i\})
&=& \prod_{i=1}^3 N^{-1}(\boldsymbol{k}\,'_i) N^{-1}(\boldsymbol{k}_i)\nonumber\\
&&\hspace{-0.2 truecm}\times\frac{1}{3}\Big\{ -(A_1A_2 +\boldsymbol{B}_1\cdot\boldsymbol{B}_2)(iB_{3x}+B_{3y})
\nonumber \\
& & \ {}\quad +
2\boldsymbol B_1\cdot\boldsymbol B_3(iB_{2x}+B_{2y})
+2\boldsymbol B_2\cdot\boldsymbol B_3(iB_{1x}+B_{1y})
\nonumber \\
& & \ {}\quad +2i\Big [A_3A_1(iB_{2x}+B_{2y})+A_3A_2(iB_{1x}+B_{1y})
\Big]
\Big\},\label{eq:x11_flip}
\end{eqnarray}
where
\begin{eqnarray}
A_i &=& (m+ x'_iM'_0)(m+ x_i  M_0) + k'^y_i k^y_i + k'^x_i k^x_i,
\nn\\
B_{i,x} &=& - (m+ x'_iM'_0) k^y_i + (m+ x_i  M_0) k'^y_i,
\nn\\
B_{i,y}& =& (m+ x'_iM'_0) k^x_i -  (m+ x_i  M_0) k'^x_i,
\nn\\
B_{i,z} &=& k'^x_i k^y_i - k'^y_i k^x_i .
\label{eq:def-ab}
\end{eqnarray}

\section{Boer-Mulders function}

The calculation of Sec.~\ref{sec:sivers} can be repeated for the
Boer-Mulders function, defined from the following quark correlation function
\begin{equation}
\label{bm}
h_{1}^\perp(x,\boldsymbol k^{\, 2}_\perp)=\epsilon^{ij} k^j_\perp\frac{M}{2\boldsymbol k_\perp^{\, 2}}
\int\frac{d\xi^-d^2\boldsymbol \xi_\perp}{(2\pi)^3}e^{-i(\xi^- k^+-\boldsymbol\xi_\perp\cdot\boldsymbol k_\perp)}
\frac{1}{2}\sum_\Lambda
\langle P\Lambda|\bar\psi(\xi^-,\boldsymbol \xi_\perp){\cal L}^\dagger_\xi i\sigma^{i+}\gamma_5{\cal L}_0\psi(0)|P
\Lambda\rangle \ .
\end{equation}
Also in this case we expand the gauge link up to the next-to leading order, and
following the same method  we used in the calculation of the Sivers function,
we find  for the Boer-Mulders function
\begin{eqnarray}
&&h_{1}^{\perp\,q}(x,\boldsymbol k_\perp^{\, 2})=-g^2
\frac{k^x-i k^y}{\boldsymbol k^{\,2}_\perp}
\frac{M}{2}
\frac{1}{(2\pi)^{11}}\frac{1}{\sqrt{(2k^+)(2k^+_1)}}
\int\frac{{\rm d}k_3^+{\rm d}^2\boldsymbol k_{3\perp}}{\sqrt{(2k_3^+)(2k^+_4)}}
\int{\rm d}^2\boldsymbol q_\perp
 \nonumber\\
&&\times
\Big\{\frac{1}{\boldsymbol{q}^{\,2}_\perp}\sum_{\Lambda,\lambda_3}
\sum_f\sum_{i.j}\sum_{k,l}
T^a_{ij}T^b_{kl}\delta_{ab}
\langle P\Lambda|b^{\dagger\, q}_{i,\,\uparrow}(k_1)b^q_{j,\,\downarrow}(k)
b^{\dagger\, f}_{k,\, \lambda_3}(k_3)b^f_{l,\, \lambda_3}(k_4)
|P\Lambda\rangle\Big\},
\label{eq:bm1}
\end{eqnarray}
 where the quark momenta are  defined as $k_1=k-q$, $k_4=k_3-q$.
The above equation corresponds to the diagram of Fig.~\ref{fig1}
with $\lambda=-\lambda_1$ and $\lambda_4=\lambda_3$ for the helicity of the interacting and spectator quarks, respectively, and $\Lambda=\Lambda'$ for
the helicity of the nucleon in the initial and final states,
i.e. the helicity is conserved at the quark-gluon vertex,
while the helicity of the struck quark
flips from the initial to the final state.
Since the nucleon state has the same helicity in the initial and final state,
the quark helicity flip must be compensated by a transfer
of one unit of orbital angular momentum.

Inserting in Eq.~(\ref{eq:bm1}) the light-cone wave-function
amplitude decomposition of the nucleon state introduced  in Sec.~\ref{sect:lcwf}, one finds the following results in terms of the
amplitudes $\psi^{(i)}$
\begin{eqnarray}
h_{1}^{\perp\,q}(x,\boldsymbol k^{\, 2}_\perp)&=&\frac{2}{3}g^2M
\frac{k^x-ik^y}{{\boldsymbol k}^{\, 2}_\perp}
\int \frac{{\rm d^2}\boldsymbol q_\perp}{(2\pi)^2}\frac{1}{\boldsymbol q^{\, 2}_\perp}
\int{\rm d}x'\int{\rm d}^2\boldsymbol t\,'_\perp
\int
d[1]d[2]d[3] \sqrt{x_1 x_2 x_3}\,{\cal H}^{\perp\,q},\nonumber\\
\label{eq:bm-overlap}
\end{eqnarray}
where the function ${\cal H}^{\perp\, q}$ for the up quark is
\begin{eqnarray}
{\cal H}^{\perp\, u}&=&
-C^{(1,2)}\tilde \phi^{(3,4)}(1,2,3)+\widetilde C^{(3,4)}\phi^{(1,2)}(1,2,3)
-C^{(3,4)}\tilde\phi^{(6)}(1,2,3)\nonumber\\
&&+\widetilde C^{(6)}\phi^{(3,4)}(1,2,3)+
	\widetilde C^{(1,2)}\phi^{(5)}(1,2,3)-C^{(5)}\tilde\phi^{(1,2)}(1,2,3),
\label{eq:H-up}
\end{eqnarray}
with
\begin{eqnarray}
\tilde \phi^{(1,2)}(1,2,3)&=&\psi^{(1)}(1,2,3)+
i(k_1^xk_2^y-k_1^yk_2^x) \psi^{(2)}(1,2,3) ,\nonumber\\
\tilde \phi^{(3,4)}(1,2,3)&=&k_1^+\psi^{(3)}(1,2,3)
	+k_2^+\psi^{(4)}(1,2,3),\nonumber\\
\tilde \phi^{(6)}(1,2,3)&=&k_1^+k_3^+\psi^{(6)}(1,2,3)
		-k_1^+k_2^+\psi^{(6)}(1,3,2) .
\end{eqnarray}
In Eq.~(\ref{eq:H-up}),  the terms containing
$C^{(1,2)}$ and $C^{(3,4)}$ describe the contribution from $S$ and $P$ wave interference,
while $C^{(5)}$ and $C^{(6)}$ are associated with the $P-D$ wave interference
terms.
The explicit expression for these functions is
\begin{eqnarray}
C^{(1,2)}&=&\delta^3(k-k_2)\Big[\delta^3(t'-k_1)\Big(\phi^{(1,2)*}(\hat 1,3,2'')
+2\phi^{(1,2)*}(2'',3,\hat 1)\Big)\nonumber\\
&&
  +\delta^3(t'-k_3)\Big(\phi^{(1,2)*}(1,\hat 3,2'')
+2\phi^{(1,2)*}(2'',\hat 3,1)\Big)
\Big]\nonumber\\
&&+\delta^3(k-k_3)\Big[\delta^3(t'-k_1)\phi^{(1,2)*}(3'',2,\hat 1)
+\delta^3(t'-k_2)\phi^{(1,2)*}(3'',\hat 2,1)\Big],\nonumber
\end{eqnarray}
\begin{eqnarray}
\tilde C^{(3,4)}&=&\delta^3(k-k_1)\left[\delta^3(t'-k_2)\left(
\tilde \phi^{(3,4)*}(3,\hat 2,1'')+2\tilde\phi^{(3,4)*}(3,1'',\hat 2)\right)\right.\nonumber\\
&&+\left.\delta^3(t'-k_3)\left(
\tilde \phi^{(3,4)*}(\hat 3,2,1'')+2\tilde\phi^{(3,4)*}(\hat 3,1'',2)\right)\right]
\nonumber\\
&&+\delta^3(k-k_3)\left[\delta^3(t'-k_1)\tilde \phi^{(3,4)*}(\hat 1,3'',2)
+\delta^3(t'-k_2)\tilde\phi^{(3,4)*}(\hat 1,3'',2)
\right],\nonumber
\end{eqnarray}
\begin{eqnarray}
C^{(3,4)}&=&\delta^3(k-k_1)\left[\delta^3(t'-k_2)\phi^{(3,4)*}(1'',\hat 2,3)
  +\delta^3(t'-k_3)\phi^{(3,4)*}(1'',2,\hat 3)\right]\nonumber\\
&&+\delta^3(k-k_2)\left[\delta^3(t'-k_1)\phi^{(3,4)*}(2'',\hat 1, 3)
+\delta^3(t'-k_3)\phi^{(3,4)*}(2'',1,\hat 3)\right],\nonumber\\
&&\nonumber\\
\tilde C^{(6)}&=&\delta^3(k-k_1)
\left[\delta^3(t'-k_2)\left(\tilde\phi^{(6)*}(1'',\hat 2,3)
+\tilde\phi^{(6)*}(\hat 2,1'',3)\right)\right.\nonumber\\
&&+\left.
\delta^3(t'-k_3)\left(\tilde\phi^{(6)*}(1'',2,\hat 3)
+\tilde\phi^{(6)*}(2,1'',\hat 3)\right)\right],\nonumber
\end{eqnarray}
\begin{eqnarray}
\tilde C^{(1,2)}&=&\delta^3(k-k_1)
\left[\delta^3(t'-k_2)\tilde\phi^{(1,2)*}(\hat 2,1'',3)+
\delta^3(t'-k_3)\tilde\phi^{(1,2)*}(2,1'',\hat 3)\right]\nonumber\\
&&+\delta^3(k-k_2)
\left[\delta^3(t'-k_1)\tilde\phi^{(1,2)*}(\hat 1,2'',3)+
\delta^3(t'-k_3)\tilde\phi^{(1,2)*}(1,2'',\hat 3)\right],\nonumber\\
&&\nonumber\\
C^{(5)}&=&\delta^3(k-k_2)
\Big[\delta^3(t'-k_1)\Big(\phi^{(5)*}(\hat 1,2'',3)
+\phi^{(5)*}(2'',\hat 1,3)\Big)\nonumber\\
&&+\delta^3(t'-k_3)\Big(\phi^{(5)*}(1,2'',\hat 3)
+\phi^{(5)*}(2'',1,\hat 3)\Big)\Big].
\label{eq:C}
\end{eqnarray}
In the above equations, the complex conjugate only acts on the wave function ${\psi}^{(i)}$.

Analogously, the function ${\cal H}^{\perp}$ for the down quark is
\begin{eqnarray}
{\cal H}^{\perp\, d}&=&
D^{(1,2)}\tilde \phi^{(3,4)}(1,2,3)-\tilde D^{(3,4)}\phi^{(1,2)}(1,2,3)
+D^{(3,4)}\tilde\phi^{(6)}(1,2,3)\nonumber\\
&&-\tilde D^{(6)}\phi^{(3,4)}(1,2,3)+
	\tilde D^{(1,2)}\phi^{(5)}(1,2,3)-D^{(5)}\tilde\phi^{(1,2)}(1,2,3),
\label{eq:H-down}
\end{eqnarray}
where the $S-P$ wave interference contribution comes from
the terms proportional to $D^{(1,2)}$ and $D^{(3,4)}$, while the remaining two terms
give the contribution from the interference of $P$ and $D$ waves.
The function $D$ in Eq.~(\ref{eq:H-down}) are defined as
\begin{eqnarray}
D^{(1,2)}&=&\delta^3(k-k_3)\left[\delta^3(t'-k_1)\phi^{(1,2)*}(\hat 1,2,3'')
  +\delta^3(t'-k_2)\phi^{(1,2)*}(1,\hat 2,3'')\right],\nonumber\\
&&\nonumber\\
\tilde D^{(3,4)}&=&\delta^3(k-k_3)\left[\delta^3(t'-k_1)
\tilde \phi^{(3,4)*}(\hat 1,2,3'')
  +\delta^3(t'-k_2)\tilde\phi^{(3,4)*}(1,\hat 2,3'')\right],\nonumber\\
&&\nonumber\\D^{(3,4)}&=&\delta^3(k-k_3)\left[\delta^3(t'-k_1)\left(
  \phi^{(3,4)*}(3'',2,\hat 1)+\phi^{(3,4)*}(3'',\hat 1,2)\right)
\right.\nonumber\\
  &&+\left.\delta^3(t'-k_2)\left(\phi^{(3,4)*}(3'',\hat 2,1)+
\phi^{(3,4)*}(3'',1,\hat 2)\right)\right],\nonumber\\
&&\nonumber\\
\tilde D^{(6)}&=&\delta^3(k-k_1)
\left[\delta^3(t'-k_2)\left(\tilde\phi^{(6)*}(3,\hat 2,1'')+\tilde\phi^{(6)*}(\hat 2,3,1'')
\right)\right.\nonumber\\
&&+\left.
\delta^3(t'-k_3)\left(\tilde\phi^{(6)*}(\hat 3,2,1'')+\tilde\phi^{(6)*}(2,\hat 3,1'')
\right)\right],\nonumber\\
&&\nonumber\\
\tilde D^{(1,2)}&=&\delta^3(k-k_2)
\left[\delta^3(t'-k_1)\left(\tilde\phi^{(1,2)*}(\hat 1,2'',3)
+\tilde\phi^{(1,2)*}(3,2'',\hat 1)\right)\right.\nonumber\\
&&+\left.
\delta^3(t'-k_3)\left(\tilde\phi^{(1,2)*}(1,2'',\hat 3)
+\tilde\phi^{(1,2)*}(\hat 3,2'',1)\right)\right],\nonumber\\
&&\nonumber\\
D^{(5)}&=&\delta^3(k-k_2)
\Big[\delta^3(t'-k_1)\Big(\phi^{(5)*}(\hat 1,2'',3)
+\phi^{(5)*}(3,2'',\hat1)\Big)\nonumber\\
&&+\delta^3(t'-k_3)\Big(\phi^{(5)*}(1,2'',\hat 3)
+\phi^{(5)*}(\hat 3,2'',1)\Big)\Big].
\label{eq:D}
\end{eqnarray}
In the above equations, the complex conjugate only acts on the wave function ${\psi}^{(i)}$.

In the model for the three-quark light cone amplitudes introduced in Sec.~\ref{sect:lcwf}, we find the following explicit results
\begin{eqnarray}
&&h_{1}^{\perp\,q}(x,\boldsymbol k^{\, 2}_\perp)=\frac{2}{3}g^2M
\frac{k^x-ik^y}{{\boldsymbol k}^2_\perp}
\int \frac{{\rm d}^2\boldsymbol q_\perp}{(2\pi)^2}\frac{1}{\boldsymbol q^{\, 2}_\perp}
\int{\rm d}x'\int{\rm d}^2\boldsymbol t\,'_\perp
\int
d[1]d[2]d[3] \sqrt{x_1 x_2 x_3}\nonumber\\
&&\times \delta(x-x_3)
\delta^2(\boldsymbol k_\perp-\boldsymbol k_{3\perp})\delta(x'-x_1)
\delta^2(\boldsymbol t\, '_\perp-\boldsymbol k\, '_{1\perp})
\,\psi^*(\{x'_i\},\{\boldsymbol k'_{i\perp}\})
\,\psi(\{x_i\},\{\boldsymbol k_{i\perp}\})\nonumber\\
&&\times3\delta_{\tau_3\tau_q}\
\left\{
\delta_{\tau_q 1/2}\tilde X^{00}(\{\boldsymbol k\,'_i\},\{\boldsymbol k_i\})
+\frac{1}{3}[\delta_{\tau_q 1/2}+2\delta_{\tau_q -1/2}]
\tilde X^{11}(\{\boldsymbol k\,'_i\},\{\boldsymbol k_i\})\right\},\label{eq:boer-lcwf}
\end{eqnarray}
where the quark momenta in the final state are $(x'_3=x,\boldsymbol k\, '_{3\perp}=\boldsymbol k_{3\perp}-\boldsymbol q_\perp)$,
$(x'_1=x',\boldsymbol k\,'_1=\boldsymbol t\, '_\perp+\boldsymbol q_\perp)$, $(x'_2=x_2,\boldsymbol k\, '_{2\perp}=\boldsymbol k_{2\perp})$.
In Eq.~(\ref{eq:boer-lcwf}), the functions $\tilde X^{00}$ and $\tilde X^{11}$ are given by,
\begin{eqnarray}
\tilde X^{00}(\{\boldsymbol k\,'_i\},\{\boldsymbol k_i\})
& = &
\prod_{i=1}^3 N^{-1}(\boldsymbol{k}\,'_i) N^{-1}(\boldsymbol{k}_i)
\Big[
(A_1A_2 +
\boldsymbol{B}_1\cdot\boldsymbol{B}_2)\tilde A_{3}\Big] ,
\label{eq:x00_tilde_noflip}\\
\tilde X^{11}(\{\boldsymbol{k}\,'\},\{\boldsymbol{k}\})\Big)
& = &
\prod_{i=1}^3 N^{-1}(\{\boldsymbol{k}\,'_i\}) N^{-1}(\{\boldsymbol{k}_i\})\nonumber\\
&&\hspace{-0.2 truecm}\times\frac{1}{3}
\Big[  (3 A_1A_2 -\boldsymbol{B}_1\cdot\boldsymbol{B}_2 )\tilde A_3
+ 2 (A_1 B_{2,x}+A_2 B_{1,x})\tilde B_{3,x}\nonumber\\
& & \quad + 2(A_1 B_{2,y}+A_2 B_{1,y})\tilde B_{3,y}
+2(A_1 B_{2,z}+A_2 B_{1,z})\tilde B_{3,z}\Big],
\label{eq:x11_tilde_noflip}
\end{eqnarray}
where the functions $A_i$ and $\boldsymbol{B}_i$ are defined in Eq.~(\ref{eq:def-ab}), and
\begin{eqnarray}
\tilde A_3&=&  (m+ x_3  M_0)(k'^x_3 +i k'^y_3)- (m+ x'_3M'_0)(k^x_3+i k^y_3),\nn
\\
\tilde B_3^x &=&  -i(m+ x'_3M'_0) (m+ x_3  M_0)+i(k'^x_3+ik'^y_3)(k^x_3+ ik^y_3),\nn
\\
\tilde B_3^y &=& (m+ x'_3M'_0) (m+ x_3  M_0)+(k'_{3,x}+ik'_{3y})(k_3^x+ ik_3^y),\nn
\\
\tilde B_3^z &=&  i(m+ x'_3M'_0)(k_3^x+ik_3^y) +i(m+ x_3  M_0)(k'^x_3+ik'^y_3).
\end{eqnarray}

\section{Results and discussion}
\label{sect:results}
The formalism described in the previous sections is applied in the
following to a specific CQM, adopting a power-law form for the momentum-dependent part of the light-cone wave function, i.e.\,
\bea
\psi(\{x_i,\boldsymbol{ k}_{i\perp }\})=
\frac{N'}{(M_0^2+\beta^2)^\gamma},
\label{eq:30}
\eea
with $N'$  a normalization factor. In Eq.~(\ref{eq:30}),
the scale $\beta$, the parameter $\gamma$ for the power-law
behaviour, and the quark mass $m$ are
taken from Ref.~\cite{Schlumpf:94a}, i.e., $\beta=0.607 $ GeV,
$\gamma=3.4$ and $m=0.267$ GeV. According to the analysis of
Ref.~\cite{Schlumpf:94b} these values lead to a very good
description of many baryonic properties.
The same parametrization of the momentum dependent part of
the LCWF in Eq.~(\ref{eq:30}) has been successfully applied
also in recent works for the calculation
of the electroweak properties of the nucleon~\cite{Pasquini:2007iz},
GPDs~\cite{Boffi:2002yy,Boffi:2003yj,Pasquini:2005dk,Pasquini:2006iv,Boffi:2007yc}
and T-even TMDs~\cite{Pasquini:2008ax,Boffi:2009sh}.

In order to fix the coupling constant appearing in Eqs.~(\ref{eq:sivers-overlap}) and (\ref{eq:bm-overlap}), we need to determine the hadronic scale of the model.
This is achieved in a model independent way following the prescription of Ref.~\cite{Pasquini:2004gc}, by matching the value of the momentum fraction carried by the valence quarks, as  computed in the model, with that obtained evolving backward the value experimentally determined at large $Q^2$. The strong coupling constant $\alpha_S(Q^2)$ entering  the evolution code  at NLO is computed by solving the
 NLO transcendental equation numerically,
\be
\ln {Q^2\over\Lambda_{\rm NLO}^2}-{4\,\pi\over\beta_0\,\alpha_s} +
{\beta_1\over\beta_0^2}\,\ln\left[
{4\,\pi\over\beta_0\,\alpha_s} + {\beta_1\over\beta_0^2}\right] = 0\,,
\label{0:10}
\ee
as obtained from the renormalization group analysis~\cite{Pasquini:2004gc,Weigl:1995hx}. It differs from the more familiar expression
\begin{equation}
{\alpha_s(Q^2) \over 4\pi}={1 \over \beta_0\ln(Q^2/\Lambda_{\rm NLO}^2)}
\left(1-{\beta_1 \over \beta_0^2}\,{\ln\ln(Q^2/\Lambda_{\rm NLO}^2)\over
\ln(Q^2/\Lambda_{\rm NLO}^2)}\right),
\label{1:10}
\end{equation}
valid only in the limit $Q^2\gg\Lambda_{\rm NLO}^2$,
 where $\Lambda_{\rm NLO}$ is the so-called QCD scale parameter.
The hadronic scale, $\mu_0^2$, consistent with the presence of valence
degrees of freedom only  is $\mu_0^2 = 0.094$ GeV$^2$, with
$\Lambda_{\rm NLO}= 0.248$ GeV.
This corresponds to a value of the strong coupling constant in Eq.~(\ref{0:10})
$\alpha_S(\mu^2_0)/(4\pi)=g^2/(4\pi)^2=0.14$, and is consistent  with the  analysis
of Refs.~\cite{Courtoy:2009pc,Courtoy:2008vi,Courtoy:2008dn} where a similar procedure was adopted.

\begin{figure}[t]
\begin{center}
\epsfig{file=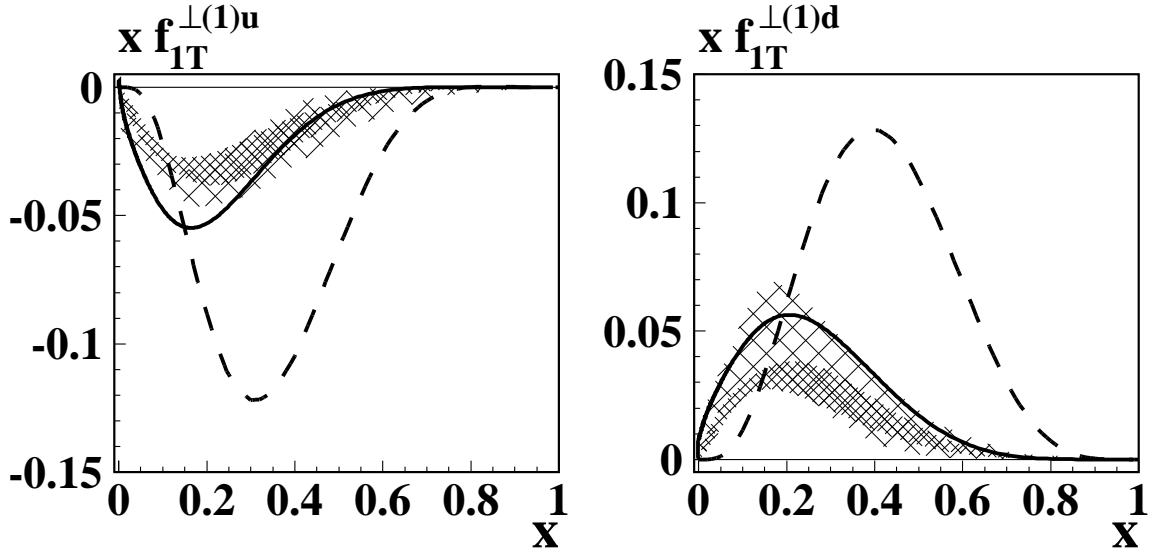,  width=\columnwidth}
\end{center}
\caption{Results for the first transverse-momentum moment of the Sivers function,
for up (left) and down (right) quark, as function of $x$. The dashed curves show the results at the hadronic scale of the model $\mu_0^2=0.094$ GeV$^2$, and the solid curves correspond to the results after NLO evolution  to $Q^2=2.5$ GeV$^2$, using the evolution pattern of the unpolarized
 parton distribution.
The lighter and darker shaded areas are the uncertainty bands due to the statistical error of the parametrizations of Ref.~\cite{Anselmino:2008sga} and Ref.~\cite{Collins:2005ie,Efremov:2004tp}, respectively.
Both parametrizations refer to an average  scale of $Q^2=2.5$ GeV$^2$.
}
\label{fig2}
\end{figure}

\begin{figure}[t]
\begin{center}
\epsfig{file=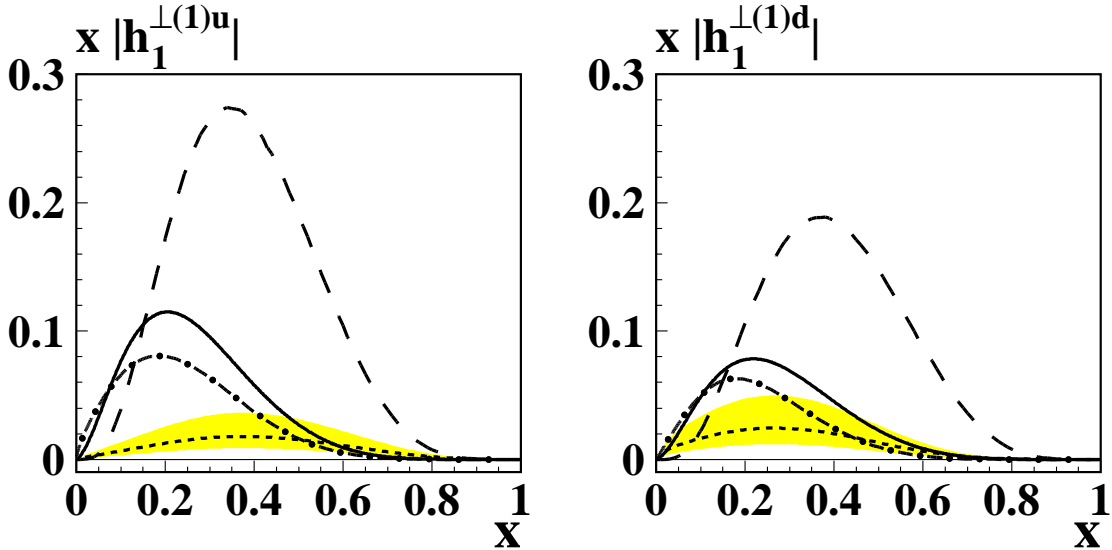,  width=\columnwidth}
\end{center}
\caption{Results for the first transverse-momentum moment of the Boer-Mulders function,
for up (left) and down (right) quark, as function of $x$.
The dashed curves show the results at the hadronic scale of the model $\mu_0^2=0.094$ GeV$^2$ and the solid curves correspond to the results after NLO evolution to $Q^2=2.4$ GeV$^2$, using the evolution pattern of the transversity distribution.
The dashed-dotted curves are the results of the phenomenological parametrization
of Refs.~\cite{Barone:2008tn,Barone:2009hw}
at the average scale of $Q^2=2.4 $ GeV$^2$, and the short-dashed curves
correspond the results  of
Refs.~\cite{Zhang:2008ez,Lu:2009ip} valid at $Q^2\approx 1$ GeV$^2$, with the shaded area describing the variation ranges allowed by positivity bounds.
}
\label{fig3}
\end{figure}

The first transverse-momentum moments of the Sivers and Boer-Mulders functions are shown in Figs.~\ref{fig2} and
\ref{fig3}, using the definition
\begin{eqnarray}
j^{(1)}(x)= \int{\rm d}^2 \boldsymbol k_\perp\frac{\boldsymbol k^{\, 2}_\perp}{2M^2}
j(x,\boldsymbol k^{\, 2}_\perp),
\end{eqnarray}
with $j=f_{1T}^{\perp\,q}$ and $j=h_1^{\perp}$, respectively.
In the figures the dashed curves correspond to the results at the hadronic scale of the model $\mu_0^2$, while the solid curves are obtained by applying a NLO evolution to $Q^2=2.5$ GeV$^2$, assuming for the first transverse-momentum moment of the Sivers function the same anomalous dimension of the unpolarized parton distribution and
for the first transverse-momentum moment of the Boer-Mulders the evolution pattern of the chiral-odd transversity distribution.
Although these are not the exact evolution patterns,
this is the standard procedure  adopted so far in model calculations~\cite{Courtoy:2009pc,Courtoy:2008vi,Courtoy:2008dn,Bacchetta:2008af} and parametrizations~\cite{Anselmino:2008sga,Collins:2005ie,Efremov:2004tp} of the T-odd TMDs, since the exact  evolution equations
are still under study~\cite{Ceccopieri:2005zz,Cherednikov:2007tw,Kang:2008ey,Zhou:2008mz,Vogelsang:2009pj,{Braun:2009vc}} and evolution codes for these distributions are not yet available.

For the Sivers function in Fig.~\ref{fig2} we also show the results from recent
parametrizations, valid at an average scale of $Q^2= 2.5$ GeV$^2$,
 obtained from a fit to available experimental
data on transverse single spin asymmetries for pion and kaon
in semi-inclusive deep inelastic scattering.
In particular, the darker shaded area represents the uncertainty due to the statistical errors in the parametrization of Ref.~\cite{Anselmino:2008sga}, while the lighter shaded area corresponds to the same for Ref.~\cite{Collins:2005ie,Efremov:2004tp}.
The model predictions for the contribution of $u$ and $d$ quarks
are of the same order of magnitude and opposite sign, and after evolution are well compatible with
the phenomenological parametrizations. The effects of the evolution are crucial to reproduce
the position of the peak  at $x\approx 0.2$ for both the $u$ and $d$
 quark distributions, and to rescale the magnitude of the distributions
within the range of the parametrizations.

A non trivial constraint in model calculations of the Sivers function is given by the Burkardt sum rule~\cite{Burkardt:2004ur}
\begin{eqnarray}
\sum_{q=u,\, d,\, s,\, g,\cdots}\
\int{\rm d}x
\,f_{1T}^{\perp \, (1)\,q}(x,\boldsymbol k^{\, 2}_\perp)=0\ ,
\label{burk-sr}
\end{eqnarray}
which corresponds to require that the net (summed over all partons)
transverse momentum due to final-state interaction
is zero~\cite{Burkardt:2004vm}.
Restricting the sum in Eq.~(\ref{burk-sr})
over the up- and down-quark contributions,
our model calculation of the Sivers function
reproduces exactly the sum rule.

In Fig.~\ref{fig3} we compare the model results for the absolute value
of the Boer-Mulders function
 with phenomenological parametrizations obtained from recent fits to
available experimental data. In particular, the dashed-dotted curve corresponds
to the analysis of Refs.~\cite{Barone:2008tn,Barone:2009hw}
at the average scale of $Q^2=2.4 $ GeV$^2$
of the HERMES~\cite{Giordano:2009hi} and COMPASS~\cite{Kafer:2008ud,Bressan:2009eu} measurements of the $\cos2\phi $ asymmetry in SIDIS,
while the short-dashed  curve shows the results  of
Refs.~\cite{Zhang:2008ez,Lu:2009ip} valid at $Q^2\approx 1$ GeV$^2$, obtained from a fit to $pd$~\cite{Zhu:2006gx} and $pp$~\cite{Zhu:2008sj} Drell-Yan data measured by the E866/NuSea Collaboration, with the shaded area describing the variation ranges allowed by positivity bounds.
We note that the available data do not allow yet a full fit of $h_1^\perp$ with its $x$ and $\boldsymbol k^{\, 2}_\perp$ dependence and these phenomenological parametrizations
are  only first attempts to extract information on  this distribution.
Upcoming experimental SIDIS data also from JLab and plans for Drell-Yan experiments at GSI will play a crucial role to better constrain these analysis.
Our model predictions after the ``approximate'' evolution to $Q^2=2.4$ GeV$^2$ are compatible with the phenomenological analysis of SIDIS data,
reproducing both the peak position and the behaviour in $x$, while are at variance with the analysis of the Drell-Yan data.
In particular we confirm the findings of Ref.~\cite{Barone:2008tn} and the expectations  from various theoretical
analysis~\cite{Burkardt:2005hp,Burkardt:2007xm,Courtoy:2009pc,Gockeler:2006zu,Pobylitsa:2003ty,Gamberg:2007wm,Bacchetta:2008af}, predicting the same sign for
both the up and down contributions, with the $u$ component of $h_1^\perp$ larger in magnitude than the corresponding component of $f_{1T}^\perp$ and the $d$ components of   $h_1^\perp$ and $f_{1T}^\perp$ with approximately the same magnitude and opposite sign.

\begin{figure}[t!]
\epsfig{file=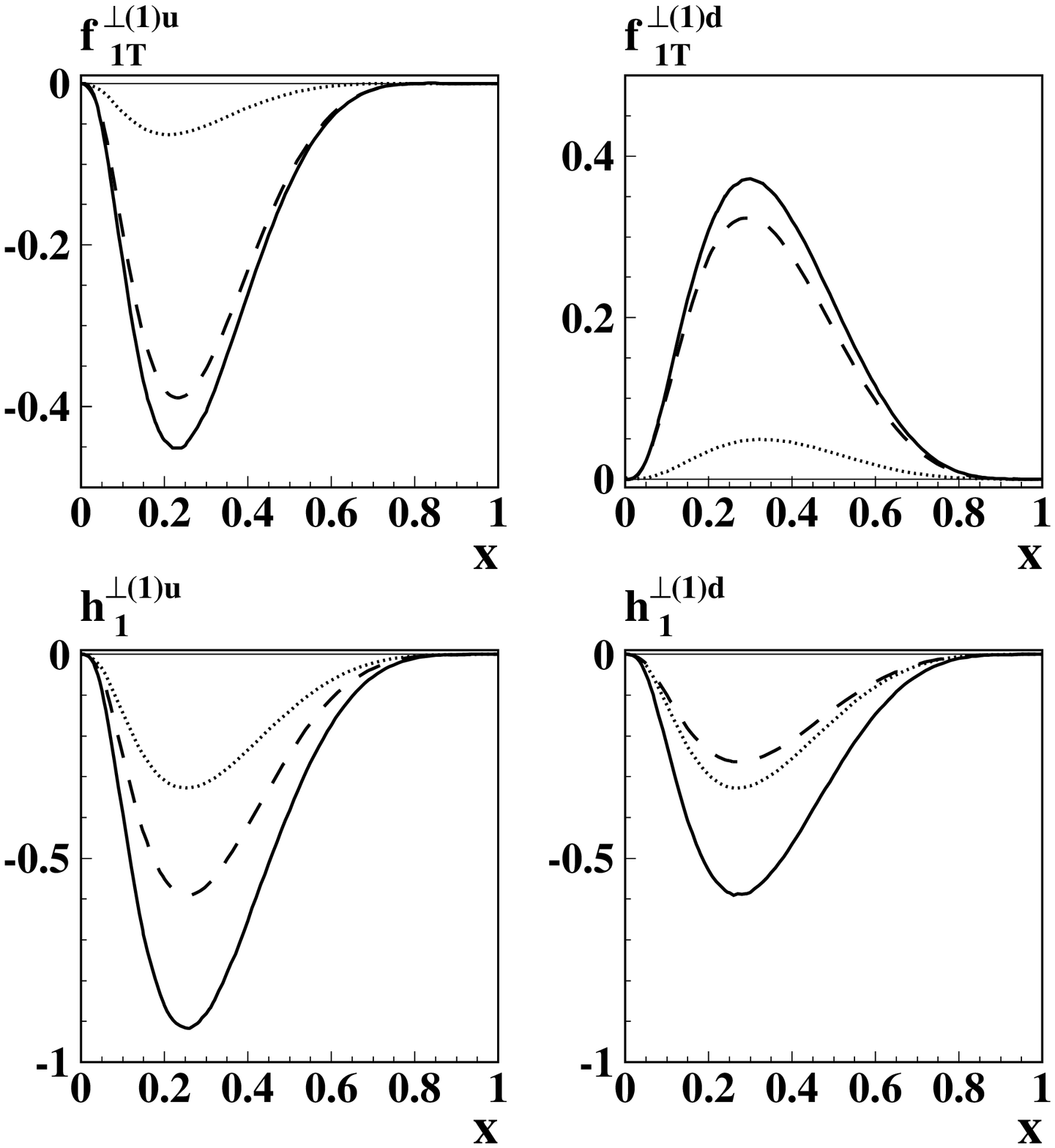,  width=0.7\columnwidth}
\caption{
Angular momentum decomposition of the first $\boldsymbol k_\perp$ moment of the Sivers function for the up (left panel ) and down (right panel) quark.
The dashed curves show the contribution from the interference of $S$ and $P$ waves,
and the dotted curves correspond to the contribution from the interference of
$P$ and $D$ waves.
The solid curves are the total results, sum of all the partial-wave contributions.}
\label{fig4}
\end{figure}
In Fig.~\ref{fig4} we show the decomposition of the Sivers and Boer-Mulders functions in the contributions from the different partial-wave amplitudes
of the nucleon LCWF.
The dashed curves correspond to the results from the interference of $S$ and $P$ waves, the dotted curves show the contribution from $P-D$ wave interference, and the solid curves are the total results, sum of all the partial wave contributions.
\begin{figure}
\epsfig{file=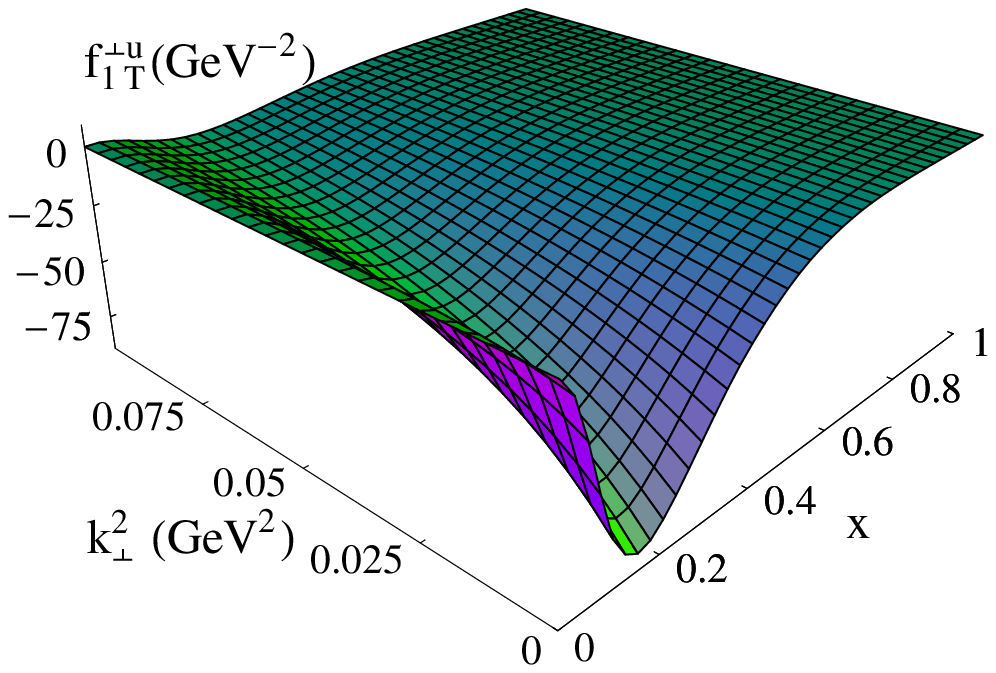,  width=0.45\columnwidth}
\hspace{0.3 truecm}
\epsfig{file=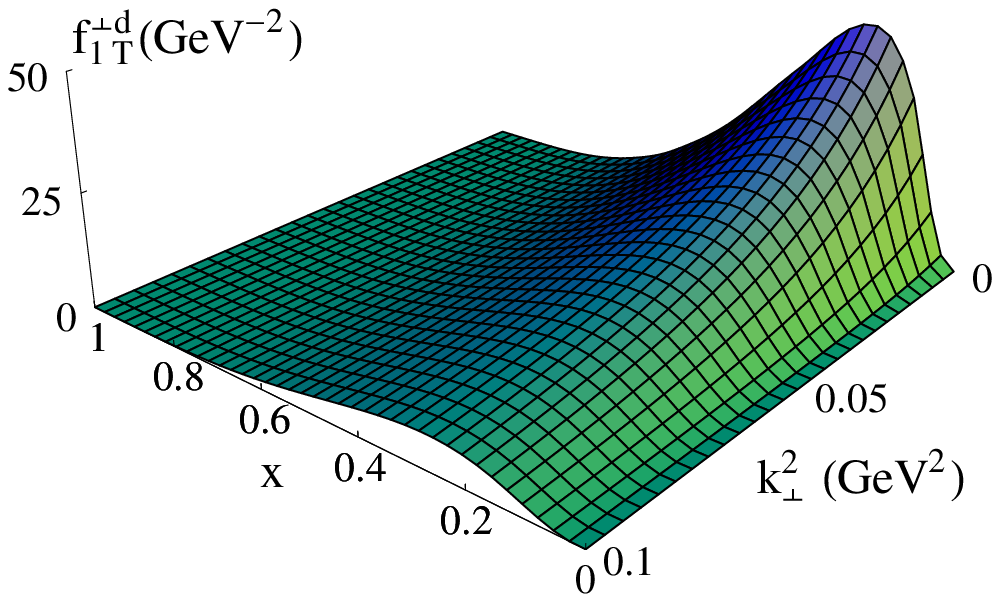,  width=0.45\columnwidth}
\caption{
The Sivers function
$f_{1T}^\perp$ as function of $x$ and $\boldsymbol k^2_\perp$ for up (left panel) and down quark (right panel).}
\label{fig5}
\end{figure}
\begin{figure}
\epsfig{file=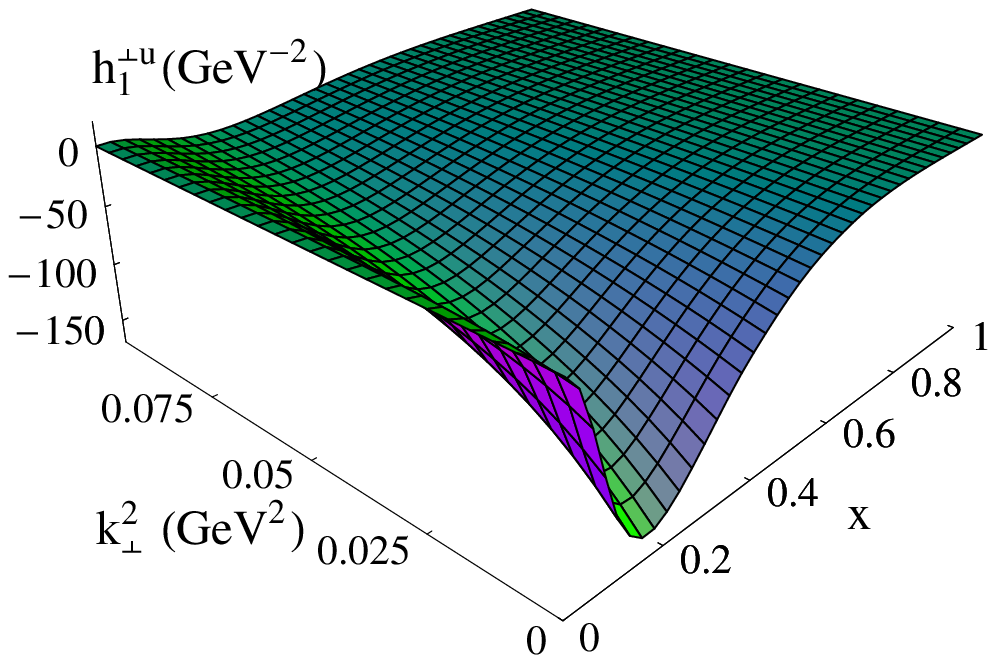,  width=0.45\columnwidth}
\hspace{0.3 truecm}
\epsfig{file=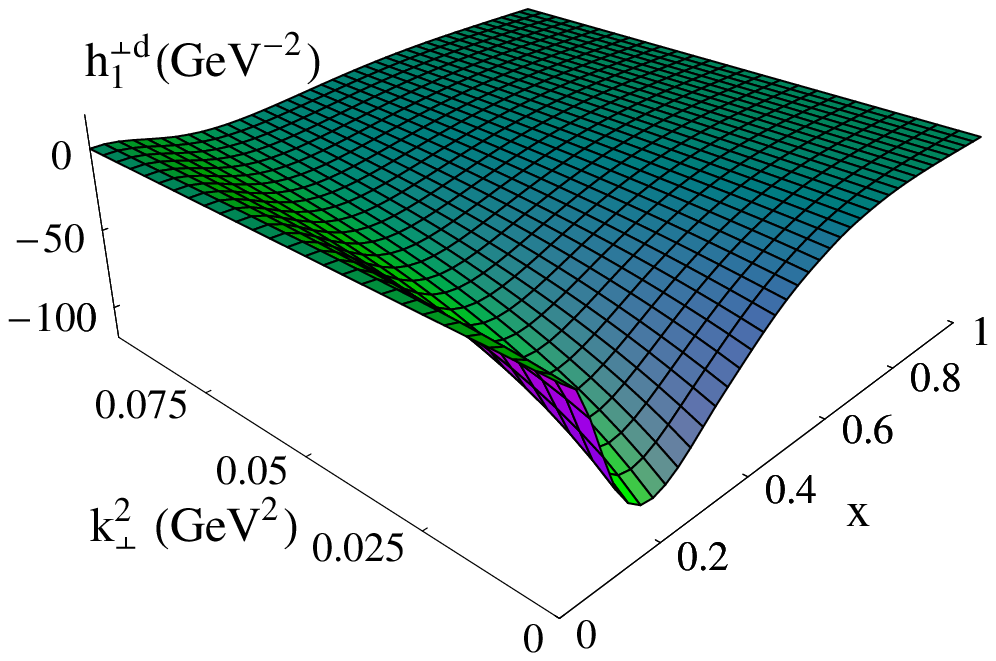,  width=0.45\columnwidth}
\caption{
The Boer-Mulders function $h_1^\perp$  as function of $x$ and $\boldsymbol k^2_\perp$ for up (left panel) and down quark (right panel).}
\label{fig6}
\end{figure}
The $S-P$ wave interference terms give the dominant contribution to the Sivers function of both $u$ and $d$ quarks, while the $P-D$ wave interference terms contribute at most by 20$\%$ of the total results.
On the other side, the relative weight of the $P-D$ wave interference terms
increases in the case of the Boer-Mulders function. It corresponds to $30\%$
of the total results for the up-quark distribution  and becomes the dominant contribution in the case of down quark, reaching up to
$60\%$ of the total result.
We also note that, contrary to the case of T-even TMDs~\cite{Pasquini:2008ax},
 the assumption of SU(6) symmetry  in the model
does not imply any proportionality between the T-odd distributions of
up and down quark.
As outlined in Ref.~\cite{Courtoy:2009pc},
this is due to the fact that in the case of the T-odd functions one is using a two-body operator associated with FSI, while for the T-even TMDs the proportionality results from the calculation with a  one-body operator.
In comparison with other model calculations,
our light-cone model predictions are similar in shape
but significantly different in magnitude from those in the 
non-relativistic CQM of Refs.~\cite{Courtoy:2009pc,Courtoy:2008vi,Courtoy:2008dn}.
The main differences in this calculation
can be traced back to the use of covariant quantization
and non-relativistic wave functions.
Furthermore, the quark-gluon interaction vertex is treated non relativistically.
Analogous discrepancies are evident in the comparison of our predictions
with the results of the bag-model~\cite{Yuan:2003wk,Courtoy:2008dn,Courtoy:2009pc}, although in this case the calculation is fully relativistic.
Since  the bag model uses covariant quantization,
the spin structure is worked out in terms
of canonical spin instead of light-cone helicity, and at the quark-gluon vertex
one has contributions from both  spin-flip and spin-conserving terms.
However, these terms reduce to helicity conserving terms ($\lambda_3=\lambda_4$ in the diagram of
 Fig.~\ref{fig1}) when written in terms of light-cone helicity, in agreement
with our model calculation.
For a more detailed discussion
on the relation between the  structure of TMDs
in terms of  canonical spin and light-cone helicity we refer to Ref.~\cite{cedric}.
\newline
\noindent
Finally, with respect to the diquark models of Refs.~\cite{Gamberg:2007wm,Bacchetta:2008af}
we have different magnitude and shape for both the Sivers and Boer Mulders functions.
The different magnitude might be due
 to the choice of different values
for the quark-gluon coupling constant as well as to the absence of D-wave components in these calculations.
Note however that our results are at variance with the calculations in the
diquark models also for  the relative magnitude between up- and down-quark
contributions.

The dependence on $x$ and $\boldsymbol k^{\, 2}_\perp$ of the Sivers and Boer-Mulders functions is shown in Figs.~\ref{fig5} and ~\ref{fig6}, respectively, for the separate up (left) and down (right) quark contributions.
The behaviour in $\boldsymbol k^2_\perp$ is very similar for the two distributions,
and does not  depend  on the quark flavour.
The  $\boldsymbol k^{\, 2}_\perp$-dependence shown in Figs.~\ref{fig5} and \ref{fig6}
is definitely not of Gaussian form.
However, following the exercise performed in Ref.~\cite{Boffi:2009sh}
 for the T-even distributions, it is interesting to compare the
model predictions for the mean square transverse momenta with the results
of the Gaussian model.
We define the mean transverse momenta ($n=1$) and the mean square transverse momenta ($n=2$) for the TMD $j(x,\boldsymbol k^{\, 2}_\perp)$ as follows
\be\label{Eq:define-mean-pT}
       \langle k_{\perp,j}^n\rangle = \frac{\int{\rm d} x\int{\rm d}^2 \boldsymbol k_\perp \;k_\perp^n\,j(x,\boldsymbol k^{\, 2}_\perp)}
                           {\int{\rm d} x\int{\rm d}^2\boldsymbol k_\perp \;j(x,\boldsymbol k^{\, 2}_\perp)} \;,
\ee
where $k_\perp=|\boldsymbol k_\perp|$.
The corresponding results for the T-odd distributions are shown in Table~I.
In the Gaussian model the following relation holds
\be\label{Eq:Gauss-relation-pT-pT2}
       \la k_\perp^2\ra \stackrel{\rm Gauss}{=} \frac{4}{\pi}\,\la k_\perp\ra^2\;,
\ee
which implies that the ratio shown in the last column of Table~I should be equal to one.
The model results deviates from unit by 10$\%$.
We also note that
the mean transverse momenta in Table~I are quite small,
much smaller than expected from phenomenological studies.
This is due to the low scale of the model, and Sudakov effects are expected to make
the $\boldsymbol k^{\, 2}_\perp$ distributions larger when evolving to larger and experimentally relevant scale.

\begin{figure}[t]
\begin{center}
\epsfig{file=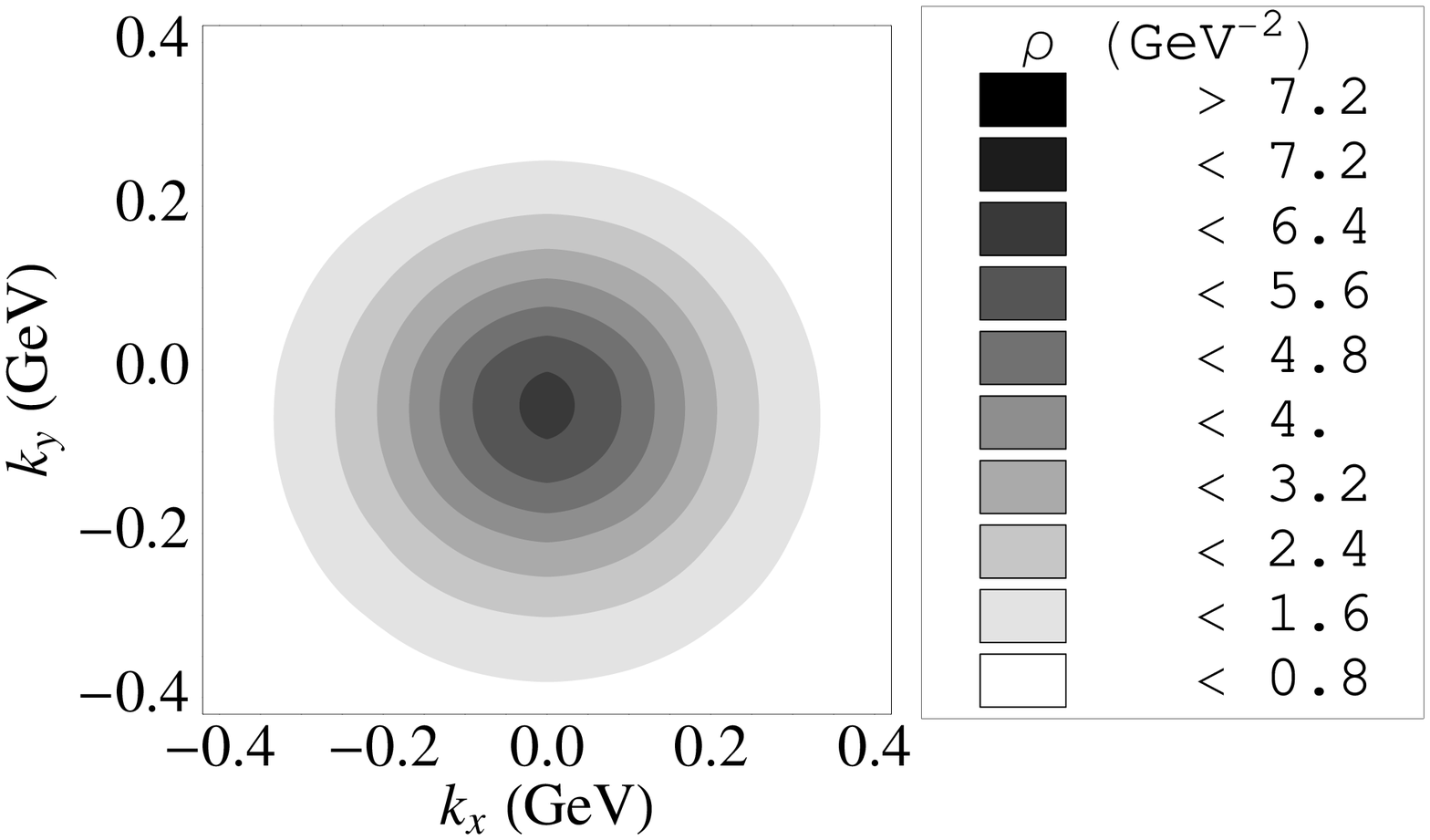,  width=0.45\columnwidth}
\hspace{0.3 truecm}
\epsfig{file=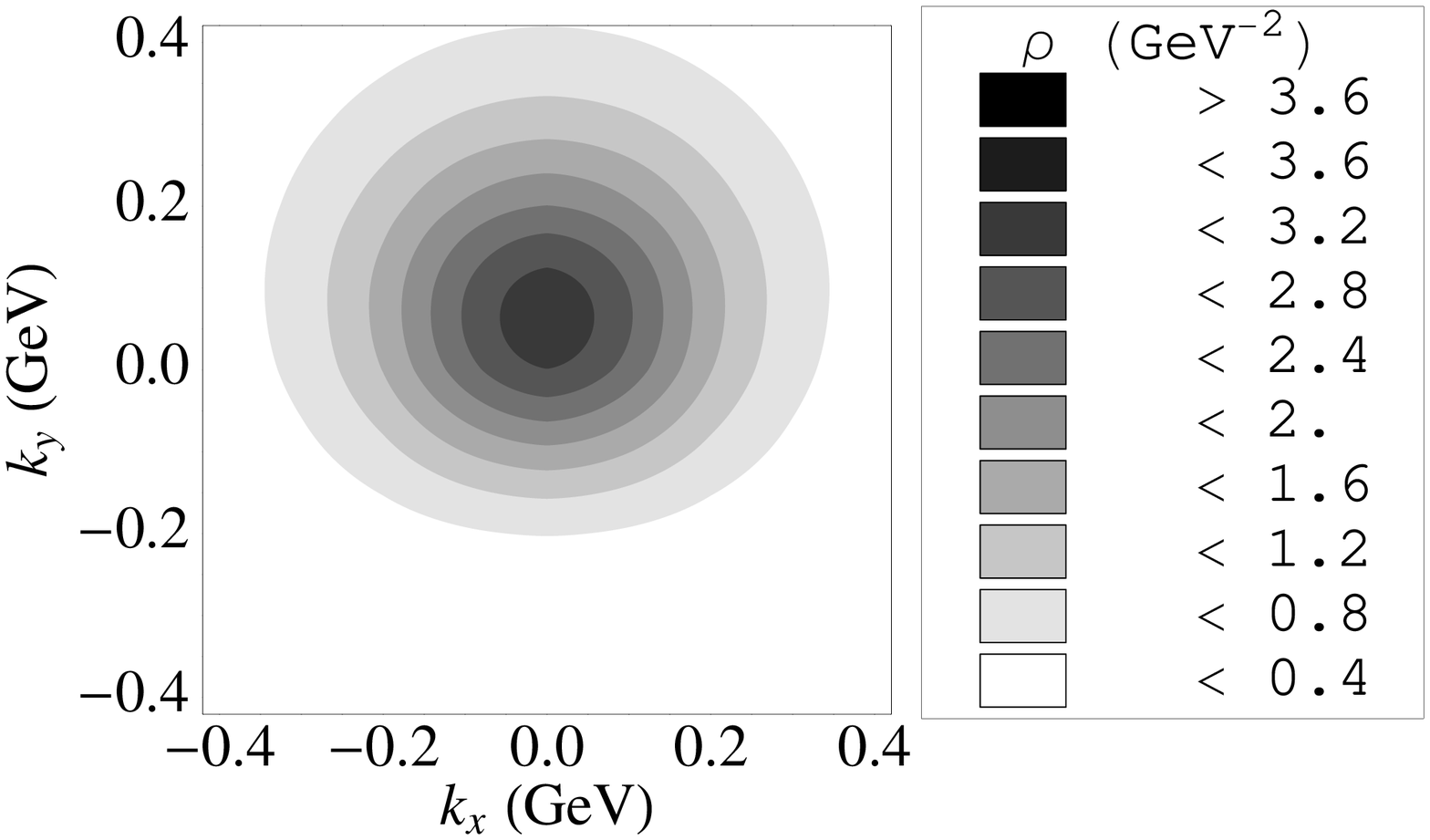, width=0.45\columnwidth}
\vspace{-2 truecm}
\end{center}
\caption{
Spin density in the transverse-momentum plane
for unpolarized quarks in a transversely polarized nucleon.
The left panel is for up quark, and the right panel for down quark.}
\label{fig7}
\end{figure}

\begin{table}[t]
\begin{tabular}{c|cc|cc|cc}
  \ \hspace{2cm} \
& \ \hspace{2cm} \
& \ \hspace{2cm} \
& \ \hspace{2cm} \
& \ \hspace{2cm} \
& \ \hspace{2cm} \
& \ \hspace{2cm} \  \cr
TMD $j$ & \multicolumn{2}{c|}{$\begin{array}{l}\langle k_\perp\rangle
  \, {\rm in \; GeV}   \end{array}$}
&  \multicolumn{2}{c|}{$\begin{array}{l}\la k_\perp^2\ra \, {\rm in \; GeV}^2 \end{array}$}
&  \multicolumn{2}{c}
{ $\displaystyle\frac{4\la k_\perp\ra^2}{\pi\la k_\perp^2\ra}$ }\cr
& & & & & &\cr
\hline
\hline
& up & down &up &down &up &down\\
$f_{1T}^\perp$&0.22&0.24&0.071&0.084&0.90&0.90\\
$h_{1}^\perp$&0.23&0.24&0.077&0.080&0.90&0.91
\end{tabular}
\label{exper}
\caption{
The mean transverse momenta and the mean square transverse
momenta of T-odd TMDs, as defined in Eq.~(\ref{Eq:define-mean-pT}),
from the light-cone CQM.
If the transverse momenta in the TMDs were Gaussian, then the
result for the ratio in the third column would be unity, see text.}
\end{table}
In Fig.~\ref{fig7}, we show the spin density in the transverse-momentum space
of unpolarized up (left panel) and down (right panel) quark
in a transversely polarized nucleon,
defined as
\begin{eqnarray}
\rho^q_{f_{1T}^\perp}(\boldsymbol k_\perp)=
\int {\rm d} x \frac{1}{2}\left[f^q_1(x,\boldsymbol k^{\, 2}_\perp)
+S^i_\perp\epsilon^{ij}k^j\frac{1}{M}f_{1T}^{\perp\,q}(x,\boldsymbol k^{\,2}_\perp)
\right]
\end{eqnarray}
with $\boldsymbol S_\perp$ the nucleon transverse-polarization vector, and $f^q_1(x,k_\perp^2)$
 the monopole distribution corresponding to spin densities for unpolarized quarks in an unpolarized target.
When $\boldsymbol S_\perp$ points in the $\hat x$ direction,
 the dipole contribution related to the Sivers function
introduces a large distortion on the monopole term,
perpendicular to both the spin and the momentum of the proton and with opposite sign for up and down quarks.
The corresponding average transverse-momentum shift is defined as
\begin{eqnarray}
\langle k^y\rangle^q_{f_{1T}^\perp}
=\frac{\int{\rm d}^2\boldsymbol k_\perp k^y \rho^q_{f_{1T}^\perp}(\boldsymbol k_\perp)}{\int{\rm d}^2\boldsymbol k_\perp \rho^q_{f_{1T}^\perp}(\boldsymbol k_\perp)}
\end{eqnarray}
and results
\begin{eqnarray}
\langle k^y\rangle^u_{f_{1T}^\perp}=\frac{M}{2}
\int{\rm d}x f_{1T}^{(1)\perp\, u}(x)=-70.31\, \mbox{MeV},
\quad\langle k^y\rangle^d_{f_{1T}^\perp}=M \int{\rm d}x f_{1T}^{(1)\perp\, d}(x)=140.62
\, \mbox{MeV}.\nonumber\\
\end{eqnarray}

The fact that the absolute value of the average transverse momentum
induced by the Sivers function is twice as large for $d$ quark than for $u$  quark is just a consequence of the Burkardt sum rule in Eq.~(\ref{burk-sr}).
This intrinsic $\boldsymbol k_\perp$ shift is the analogous
of the dipole deformation related to the GPD $E$ in impact-parameter
space~\cite{Meissner:2009ww,Diehl:2005jf}.
Although it is not possible
to establish a model independent relation between the Sivers function
and the GPD $E$~\cite{Meissner:2009ww},
we note that the LCWF overlap representation of $E$, for vanishing longitudinal momentum transfer, is given in terms of the same combinations
of light-cone amplitudes  parametrizing
the Sivers function in the one gluon-exchange approximation, but evaluated for different values of quark
variables~\cite{Ji:2002xn,Boffi:2002yy}.
The values for the average shifts in impact-parameter space
within the present light-cone quark model
were found $\langle b^y\rangle^u=\kappa^u/(2M)=0.20\, \mbox{fm}$
and $\langle b^y\rangle^d=\kappa^d/(M)=-0.33 \, \mbox{fm}$~\cite{Pasquini:2007xz}, where $\kappa^q$ is the quark contribution to the proton anomalous magnetic moment.

Analogously,  the spin density of
transversely polarized quarks
and unpolarized nucleon is related to the Boer-Mulders effect by
\begin{eqnarray}
\rho^q_{h_{1}^\perp}(\boldsymbol k_\perp,\boldsymbol s_\perp)=
\int {\rm d} x \frac{1}{2}\left[f_1^q(x,\boldsymbol k^{\, 2}_\perp)
+s^i\epsilon^{ij}k^j\frac{1}{M}h_1^{q\,\perp}(x,k_\perp^2)\right],
\end{eqnarray}
where $\boldsymbol s_\perp$ is the quark transverse-polarization vector.
\begin{figure}[t]
\begin{center}
\epsfig{file=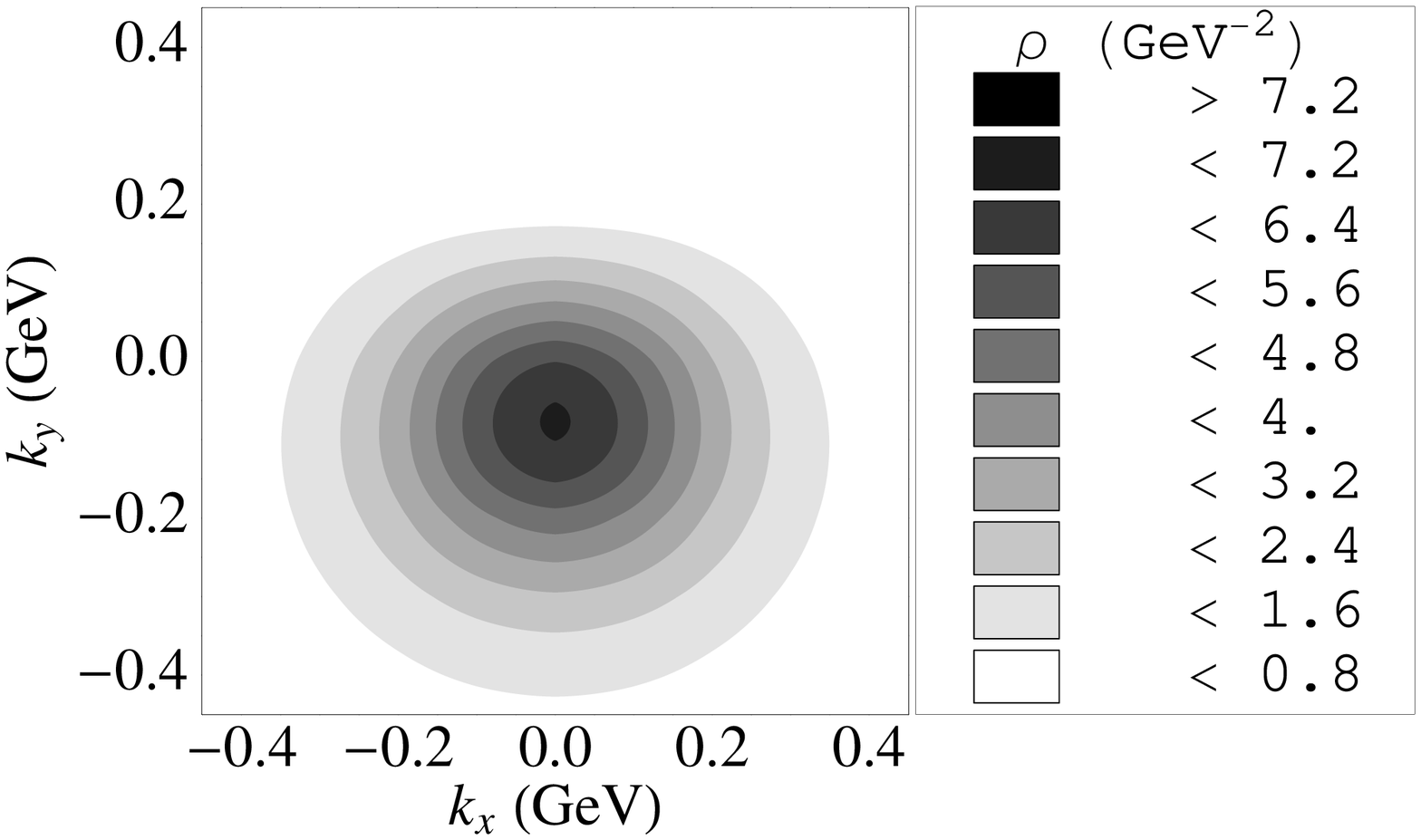,  width=0.45\columnwidth}
\hspace{0.3 truecm}
\epsfig{file=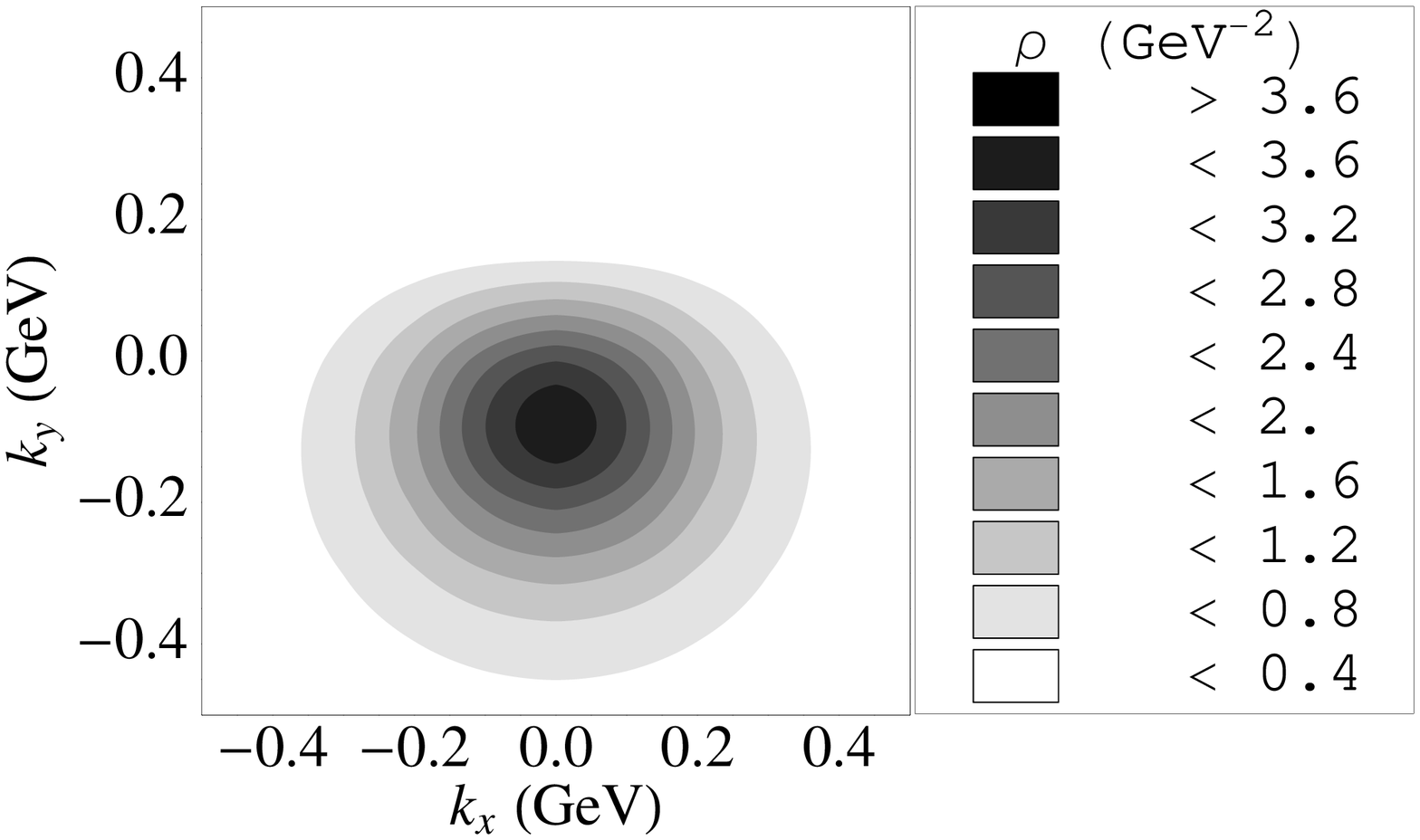, width=0.45\columnwidth}
\vspace{-2 truecm}
\end{center}
\caption{
Spin density in the transverse-momentum plane
for transversely polarized quarks in an unpolarized nucleon.
The left panel is for up quark, and the right panel for down quark.}
\label{fig8}
\end{figure}
In Fig.~\ref{fig8} we show the spin density for quark polarization in the $\hat x$ direction.
Since the Boer-Mulders function is negative for both up and down quarks,
the sideway shift is always in the positive $\hat y$ direction.
The corresponding average dipole distortion is
\begin{eqnarray}
\langle k^y\rangle^u_{h_{1}^\perp} =\frac{M}{2}\int{\rm d}x h_{1}^{(1)\perp\, u}(x)=-159.40 \, \mbox{MeV},\quad
\langle k^y\rangle^d_{h_{1}^\perp} =M \int{\rm d}x h_{1}^{(1)\perp\, d}(x)=-215.73 \, \mbox{MeV}.
\nonumber\\
\end{eqnarray}
Although the  Boer-Mulders function is smaller in magnitude for down quark than for
up quark, one observes that the average sideways distortion for down quark is
stronger.
This is because the monopole
distribution related to $f_1^q$ is twice as large  for up quarks as for down quarks, therefore adding the dipole contributions results in a more pronounced distortion for down quarks than for up quarks.
The corresponding dipole distribution in impact-parameter space is
described by the chiral odd GPDs $E_T+2\tilde H_T$.
As found in Ref.~\cite{Pasquini:2005dk}, these GPDs
for zero longitudinal momentum transfer are given by
the same combination of LCWFs which enter the calculation of 
$h_{1}^\perp$ in the one-gluon exchange approximation,
but at different kinematics.
The corresponding average distortion in impact-parameter space
is proportional to the tensor anomalous magnetic moment $\kappa^q_T$,
and in the present light-cone quark model is
given by $\kappa^u_T/(2M)=0.42 \,\mbox{fm}$
and $\kappa^d_T/(M)=0.55 \,\mbox{fm}$
for up and down quark, respectively~\cite{Pasquini:2007xz}.

\section{Conclusions}
In this paper we have investigated the naive-time-reversal-odd
quark distributions, the quark Sivers and Boer-Mulders functions,
in a light-cone quark model. The final-state interaction
effects are calculated by approximating the gauge link operator with a
one-gluon exchange interaction.
In this framework, we have derived
the general formalism for the T-odd quark distributions in terms
of overlap of light-cone wave function amplitudes describing the different orbital angular momentum components of the nucleon state.
This model independent expressions are particularly suitable to
emphasize the  correlations
of quark transverse momentum and transverse polarizations of the nucleon
and of the quark.
For numerical estimates, the nucleon light-cone wave-function has been
constructed by assuming a light-cone constituent quark model with SU(6) spin-flavor symmetry and a momentum-dependent part
which is spherically symmetric. Under this assumption the orbital angular momentum content of the wave function is fully generated by the Melosh rotations which boost the rest-frame spin into the light-cone.
As a result, we found explicit expressions for the light-cone amplitudes
which match the analytic structure expected from model-independent arguments~\cite{Ji:2002xn,Burkardt:2002uc,Ji:2003yj}.
The model dependence enters the choice of the momentum-dependent part of the light-cone wave function. In this work, we adopted a phenomenological description, by assuming
 a specific functional form with parameters fitted to hadronic structure constants. The same wave function was used to predict many other hadronic properties, providing a good description of available experimental data, and being able to capture the main features of the quark contribution to hadronic structure functions, like parton distributions~\cite{Pasquini:2006iv}, generalized parton distributions~\cite{Boffi:2002yy,Boffi:2003yj,Pasquini:2005dk,Pasquini:2007xz},
nucleon form factors~\cite{Pasquini:2007iz} and distribution amplitudes~\cite{Pasquini:2009ki}, and T-even transverse momentum dependent quark distributions~\cite{Pasquini:2008ax,Boffi:2009sh}.
\newline
\noindent
The corresponding results for the Sivers and Boer-Mulders function have been
presented in this paper by showing the decomposition into the contributions from different orbital angular momentum components.
Both functions require a transfer of one unit of orbital angular momentum between the initial and final states.
In particular, the Sivers function for both up and down quark is dominated by the interference of $S$- and $P$-wave components, while the $P-D$ wave interference terms contribute at most by $20 \%$ of the total results.
On the other side, the relative weight of the $P-D$ wave interference terms increases in the case of the Boer-Mulders function, in particular for the down-quark component.
Furthermore, the model results for the Sivers function satisfy exactly the so-called Burkardt sum rule, which is a non-trivial constraint for model calculations and parametrizations.

In order to compare with phenomenological parametrizations obtained from a fit
to available experimental data for semi-inclusive deep inelastic scattering and Drell-Yan processes, we evolved
the model results to the experimental scale. Since the exact evolution equations for the T-odd quark distributions are still under study, we used those evolution equations which seem most promising to be able to simulate the correct evolution.
We evolved the first transverse-momentum moment of the Sivers function by means of the evolution pattern of the unpolarized parton distribution, while for the first transverse-momentum moment of the Boer-Mulders
we used the evolution pattern of the transversity.
After evolution, the model results are consistent with the available parametrizations,
especially for the Sivers function.
There is agreement between the signs of the various flavor components, and also
for the magnitude and  the position of the maxima in $x$.
These findings encourage further phenomenological applications
of the model to describe azimuthal asymmetries in hadronic reactions.

We also found that the $x$ and $
\boldsymbol k_\perp^2$ dependence
is similar for the
Sivers and Boer-Mulders functions, and approximately independent on the quark flavor. In particular, the $\boldsymbol k_\perp^2$ is not of Gaussian form.
However,
it is worthwhile  to
evaluate   the degree of approximation introduced
by the Gaussian Ansatz within the model
in the calculation of  observables.
This task is left for future applications of the model.

Finally, we discussed the spin densities in the transverse-momentum space
related to the Sivers and Boer-Mulders effects, showing that they are consistent with the model results for the corresponding spin densities in the impact-parameter space described by  generalized parton distributions.
\acknowledgments

B.P. is grateful to A. Bacchetta, F. Conti, A. Courtoy and M. Radici for discussions, and to
the Nuclear Science Division of Lawrence Berkeley National Laboratory,
where this work was initiated, for hospitality.
This work was supported in part  by
the Research Infrastructure Integrating Activity
``Study of Strongly Interacting Matter'' (acronym HadronPhysics2, Grant
Agreement n. 227431) under the Seventh Framework Programme of the
European Community, by the Italian MIUR through the PRIN
2008EKLACK ``Structure of the nucleon: transverse momentum, transverse
spin and orbital angular momentum'',
and by the U.S. Department of Energy
under contracts DE-AC02-05CH11231 and DE-AC02-76SF00515.
We are grateful to RIKEN,
Brookhaven National Laboratory and the U.S. Department of Energy
(contract number DE-AC02-98CH10886) for providing the facilities
essential for the completion of this work.

\end{document}